\documentstyle[12pt]{article}

\advance \topmargin by -\headheight
\advance \topmargin by -\headsep

\evensidemargin \oddsidemargin
\def\ie{{\em i.e.}}
\def\ie{\hbox{\it i.e.}}

\def\CC{{\mathchoice
{\rm C\mkern-8mu\vrule height1.45ex depth-.05ex
width.05em\mkern9mu\kern-.05em}
{\rm C\mkern-8mu\vrule height1.45ex depth-.05ex
width.05em\mkern9mu\kern-.05em}
{\rm C\mkern-8mu\vrule height1ex depth-.07ex
width.035em\mkern9mu\kern-.035em}
{\rm C\mkern-8mu\vrule height.65ex depth-.1ex
width.025em\mkern8mu\kern-.025em}}}

\def\RR{{\rm I\kern-1.6pt {\rm R}}}

\def\ZZ{{\rm Z}\kern-3.8pt {\rm Z} \kern2pt}
\def\IB{\relax{\rm I\kern-.18em B}}
\def\ID{\relax{\rm I\kern-.18em D}}
\def\II{\relax{\rm I\kern-.18em I}}
\def\IP{\relax{\rm I\kern-.18em P}}

\def\np{Nucl. Phys.}
\def\pl{Phys. Lett.}

\def\jhep{J. High Energy Phys.}

\newcommand{\beq}{\begin{equation}}
\newcommand{\eeq}{\end{equation}}
\newcommand{\rc}{\nonumber\\}
\newcommand{\bear}{\begin{eqnarray}}
\newcommand{\eear}{\end{eqnarray}}

\def\to{\rightarrow}

\def\to{\rightarrow}

\newfont{\namefont}{cmr10}
\newfont{\addfont}{cmti7 scaled 1440}
\newfont{\boldmathfont}{cmbx10}
\newfont{\headfontb}{cmbx10 scaled 1728}
\renewcommand{\theequation}{{\rm\thesection.\arabic{equation}}}
\begin{document}
\begin{titlepage}

\begin{center} \Large \bf Let's Twist Again: General Metrics of $G_2$ Holonomy
from Gauged Supergravity

\end{center}

\vskip 0.3truein
\begin{center}
J. D. Edelstein${}^{\,\sharp\,\dagger\,*}$
\footnote{jedels@math.ist.utl.pt},
A. Paredes${}^{\,*}$
\footnote{angel@fpaxp1.usc.es}
and
A.V. Ramallo${}^{\,*}$
\footnote{alfonso@fpaxp1.usc.es}

\vspace{0.3in}
${}^{\sharp\,}$Departamento de Matem\'atica, Instituto Superior Tecnico \\
Av. Rovisco Pais, 1049--001, Lisboa, Portugal

\vspace{0.3in}

${}^{\dagger\,}$Instituto de F\'\i sica de La Plata -- Conicet, Universidad
Nacional de La Plata \\
C.C. 67, (1900) La Plata, Argentina

\vspace{0.3in}

${}^{\,*}$Departamento de F\'\i sica de Part\'\i culas, Universidad de
Santiago de Compostela \\
E-15782 Santiago de Compostela, Spain
\vspace{0.3in}

\end{center}
\vskip 1truein

\begin{center}
\bf ABSTRACT
\end{center}

We construct all complete metrics of cohomogeneity one
$G_2$ holonomy with $S^3\times S^3$ principal orbits from gauged
supergravity. Our approach rests on a generalization of the
twisting procedure used in this framework. It corresponds to a
non-trivial embedding of the special Lagrangian three--cycle
wrapped by the D6--branes in the lower dimensional supergravity.
There are constraints that neatly reduce the general ansatz to
a six functions one. Within this approach, the Hitchin system
and the flop transformation are nicely realized in eight
dimensional gauged supergravity.

\vskip2.6truecm
\leftline{US-FT-5/02\hfill November 2002}
\leftline{hep-th/0211203}
\smallskip
\end{titlepage}
\setcounter{footnote}{0}


\setcounter{equation}{0}
\section{Introduction}

Four dimensional supersymmetric Yang--Mills theory arise in
M--theory on a manifold $X$ with $G_2$ holonomy. If the manifold
is large enough and smooth, the low energy description is given in
terms of a purely gravitational configuration of eleven
dimensional supergravity. The gravity/gauge theory correspondence
then allows for a geometrical approach to the study of important
aspects of the strong coupling regime of supersymmetric
Yang--Mills theory such as the existence of a mass gap
\cite{ach1,amv}, chiral symmetry breaking \cite{amv}, confinement
\cite{ach2}, gluino condensation \cite{amv,vafa}, domain walls
\cite{achv} and chiral fermions \cite{achw}. These facts led, in
the last two years, to a concrete and important physical
motivation to study compact and non--compact seven-manifolds of
$G_2$ holonomy.

Up to last year, there were only three known examples of
complete metrics with $G_2$ holonomy on Riemannian manifolds
\cite{bs,gpp}. They correspond to $\RR^3$ bundles over $S^4$ or
${\CC\IP}^2$, and to an $\RR^4$ bundle over $S^3$. These manifolds
develop isolated conical singularities corresponding,
respectively, to cones on ${\CC\IP}^3$, $SU(3)/U(1)\times U(1)$,
or $S^3\times S^3$, and the dynamics of M--theory on them has been
recently studied in great detail \cite{aw}. See also \cite{tf}.
In the last case, in
particular, it was shown that there is a moduli space with three
branches, and the quotient by a finite subgroup of $SU(2)$ leads
either to the uplift of D6--branes wrapping a special Lagrangian
$S^3$ in a Calabi--Yau three-fold, or to a smooth manifold
admitting no normalizable supergravity zero modes. M--theory on
the latter has no massless fields localized in the
transverse four-dimensional spacetime. By a smooth interpolation
between these manifolds, M--theory realizes the mass gap of ${\cal
N}=1$ supersymmetric four-dimensional gauge theory
\cite{ach1,amv}; this geometric dual description corresponding,
however, to type IIA strings at infinite coupling.

We will concentrate on this paper in the case of an $\RR^4$ bundle
over $S^3$. Supersymmetry and holonomy matching indicate that a
large class of $G_2$ holonomy manifolds, describing the M--theory
lift of D6--branes wrapping a special Lagrangian three-cycle on a
Calabi--Yau three--fold, must exist \cite{jaume}. Constructing
their complete metrics is an important issue in improving our
understanding of the strongly coupled infrared dynamics of ${\cal
N}=1$ supersymmetric gauge theories. For example, a new $G_2$
holonomy manifold with an asymptotically stabilized $S^1$ --thus
describing the M--theory lift of wrapped D6--branes, mentioned in
the previous paragraph, in the case of finite string coupling--
was recently found \cite{bggg}. This solutions is asymptotically
locally conical (ALC) --near infinity it approaches a circle
bundle with fibres of constant length over a six--dimensional
cone--, as opposed to the asymptotically conical (AC) solutions
found in \cite{bs,gpp}.

There have been some attempts of a generic approach to build this
kind of complete and non--singular metrics of $G_2$ holonomy. A
rather general system of first--order equations for the metric was
obtained in \cite{CGLP1,CGLP2} from the BPS domain wall
equations corresponding to an auxiliary superpotential. Three
types of regular metrics were shown to arise from this system, in
which the orbits degenerate respectively to $S^3$ \cite{bggg},
$T^{1,1}$ \cite{CGLP2} and $S^2$ \cite{CGLP3}. With the notable
exception of the first one, the solutions are only known numerically.
A different fairly general technique was used in \cite{bran} by
directly exploiting the fact that $G_2$ holonomy metrics are
determined by a three--form that is closed and co-closed. 
Following a different approach, Hitchin gave a prescription dealing
with a Hamiltonian system, which is obtained by extremising
diffeomorphism invariant functionals on certain differential
forms, that leads to metrics of $G_2$ holonomy \cite{hit}. This
procedure was then exploited \cite{Chong} to obtain a general
system of first--order equations for metrics of $G_2$ holonomy with
$S^3\times S^3$ principal orbits that was shown to encompass
the previous ones. It was also shown in \cite{Chong}
that, through different contraction limits, $G_2$ metrics with
$S^3\times T^3$ principal orbits can be attained \cite{gyz}.

On the other hand, a more systematical approach, started in \cite{mn},
explicitly exploits the fact that these metrics come from the uplift
of D6--branes wrapping special Lagrangian three--cycles on a Calabi--Yau
three--fold. The key issue is given by the non--trivial geometry of the
world--volume that forces supersymmetry to be appropriately twisted such
that covariantly constant Killing spinors are supported \cite{bvs}. A
natural framework to perform the above mentioned twisting is given by
lower dimensional gauged supergravities, whose domain wall solutions
usually correspond to the near horizon limit of D--brane configurations
\cite{bst} thus giving directly the gravity dual description of the
gauge theories living on their world--volumes. This approach has been
largely followed throughout the literature on the subject
\cite{mn2}--\cite{muchmore}. In particular, a generic approach to
obtain $G_2$ holonomy manifolds from eight--dimensional gauged
supergravity was undertaken in \cite{hs1}, where it was shown that
the conventional twisting prescription should be generalized in a way
that involves non-trivially the scalar fields that arise in lower
dimensional gauged supergravity.

There is a wide spread believe that the gauged supergravity approach
is quite limited to a subset of solutions whose asymptotics is related
to near horizon geometries of D--branes. In the particular case of
our interest, it is well--known that the D6--brane solution is viewed
in eleven dimensions as a Taub-NUT space whereas its near horizon limit
is described by an Eguchi--Hanson metric. The former goes asymptotically
to $\RR^3\times S^1$ as opposed to the latter that goes as $\RR^4$. It
is then somehow unexpected to find solutions corresponding to ALC $G_2$
manifolds in lower dimensional gauged supergravities. Another argument
in this line is the following. There is a flop transformation in
manifolds of $G_2$ holonomy with $S^3\times S^3$ principal orbits that
interchanges the fibre $S^3$ with the base one. This operation, from the
point of view of eight--dimensional gauged supergravity, would amount
to an exchange between the external sphere and the one where the
D6--brane is wrapped. Then, there seems to be no room for the flop within
the gauged supergravity approach. So, in particular, flop invariant
solutions as the one obtained by Brandhuber, Gomis, Gubser and Gukov
\cite{bggg} would not be obtainable from gauged supergravity. In this
paper, we are going to show that this is not the case. It is possible
to further extend the twisting conditions in that framework in such a way
that general cohomogeneity one metrics of $G_2$ holonomy with principal
orbits $S^3 \times S^3$ turn out to be obtainable within the framework
of eight dimensional gauged supergravity.

The generalized twisting procedure that we propose corresponds to a
non-trivial embedding of the special Lagrangian three--cycle wrapped by
the D6--branes in the lower dimensional supergravity. It is important
to remark that we are using the name ``D6--brane'' in quite a loose
sense. Meaningly, many $G_2$ manifolds do not correspond to D6--branes
wrapping special Lagrangian three--cycles but to the uplift of resolved
conifolds with RR fluxes piercing the blown--up $S^2$. Starting from the
general ansatz, we found a set of constraints that neatly reduce it to a six
functions one. This makes connection with previous works in the
literature where six functions ansatz\"e were taken as a starting
point. The Hitchin system \cite{hit} turns out to be an elegant
general solution of the constraints. Finally, the flop transformation
becomes nicely realized in eight dimensional gauged supergravity. Then,
not surprisingly, flop invariant solutions (as that in \cite{bggg})
emerge in this formalism.

The plan for the rest of the paper is as follows. In section 2 we
perform a detailed study of the case of D6-branes on a special
Lagrangian round three sphere in a manifold with the topology of the
complex deformed conifold. We start by analyzing the possible
realizations of supersymmetry for the round ansatz and the corresponding
generalized twist. Then we formulate our results in terms of the
calibrating closed and co-closed three-form associated to manifolds of
$G_2$ holonomy. The analysis carried out for the round case reveals the
key points which must be taken into account in the more general case
studied in section 3. In this section, a general ansatz with triaxial
squashing is considered. The subsequent analysis shows that one must
impose certain algebraic constraints on the functions of the ansatz if
we require our solution to be supersymmetric.

In section 4 we demonstrate that our formalism provides a realization of the
Hitchin system. Some particular cases are studied in section 5,
including the flop invariant and the conifold--unification metrics, which
were never obtained by using eight dimensional gauged supergravity so
far. In section 6 we summarize  our results and draw some conclusions.
In appendix A we collect the relevant formulae of eight dimensional gauged
supergravity. Finally, for completeness in
appendix B a Lagrangian approach to the round case is presented.

\setcounter{equation}{0}
\section{D6-branes on the round 8d metric }
\medskip

The first case we will analyze corresponds to a D6-brane wrapping a
three--cycle
in such a way that the corresponding eight dimensional metric $ds^2_8$
contains a round three-sphere. Accordingly, we will adopt the following
ansatz for the metric:
\beq
ds^2_8\,=\,e^{2f}\,dx_{1,3}^2\,+\,
e^{2h}\,d\Omega_3^2\,+\,dr^2\,\,,
\label{uno}
\eeq
where $dx_{1,3}^2=-(dx^0)^2+(dx^1)^2+(dx^2)^2+(dx^3)^2$, $f$ and $h$ are
functions of the radial coordinate $r$ and $d\Omega_3^2$ is the metric of the
unit $S^3$. It is convenient to parametrize this three-sphere by means of a
set of left invariant one-forms $w^i$, $i=1,2,3$, of the SU(2) group manifold
satisfying:
\beq
dw^i\,=\,{1\over 2}\,\epsilon_{ijk}\,w^j\,\wedge\,w^k\,\,.
\label{dos}
\eeq
In terms of three Euler angles $\theta, \varphi$ and $\psi$,
the $w^i$'s are:
\bear
w^1&=&\cos\varphi\,d\theta\,+\,\sin\theta\sin\varphi\, d\psi\,\,,\rc
w^2&=&\sin\varphi\,d\theta\,-\,\sin\theta\cos\varphi\, d\psi\,\,,\rc
w^3&=&d\varphi\,+\,\cos\theta\, d\psi\,\,,
\label{tres}
\eear
and $d\Omega_3^2$ is:
\beq
d\Omega_3^2\,=\,{1\over 4}\,\sum_{i=1}^3\,(\,w^i\,)^2\,\,.
\label{cuatro}
\eeq

In this section we will study some supersymmetric configurations
of eight dimensional gauged supergravity \cite{ss} whose spacetime
metric is of the form displayed in eq. (\ref{uno}). The aspects of
this theory which are relevant for our analysis have been
collected in appendix A. In the configurations studied in the
present section, apart from the metric, we will only need to
excite the dilatonic scalar $\phi$ and the SU(2) gauge potential
$A_{\mu}^i$. Actually, we will require that, when uplifted to
eleven dimensions, the unwrapped part of the metric be that
corresponding to flat four dimensional Minkowski spacetime. This
condition determines the following relation between the function
$f$ and the field $\phi$: \beq f\,=\,{\phi\over 3}\,\,.
\label{cinco} \eeq (See the uplifting formulae in appendix A).

We will assume that the non-abelian gauge potential $A_{\mu}^i$ has only
non-vanishing components along the directions of the $S^3$. Actually, we will
adopt an ansatz in which this field, written as a one-form, is given by:
\beq
A^i\,=\,\Big(\,g\,-\,{1\over 2}\,\Big)\,  w^i\,\,,
\label{seis}
\eeq
with $g$ being a function of the radial coordinate $r$. Notice that  in
ref.\cite{en} the value $g=0$ has been taken. The field strength
corresponding to the potential (\ref{seis}) is:
\beq
F^i\,=\,g'\,dr\,\wedge\,w^i\,
+\,{1\over 8}\,(\,4g^2-1\,)\,\epsilon^{ijk}
\,w^j\,\wedge\,w^k\,\,.
\label{siete}
\eeq
By plugging our ansatz of eqs. (\ref{uno}), (\ref{cinco}) and (\ref{seis})
in the Lagrangian of  eight  dimensional gauged supergravity, one arrives
at an effective problem in which one can find a superpotential and the
corresponding first-order domain wall equations. This approach has been
followed in appendix B. In this section we will find this same
first-order equations by analyzing the supersymmetry transformations of
the fermionic fields. As we will verify soon, this last approach will
give us the hints we need to extend our analysis to metrics more general
than the one written in eq. (\ref{uno}).

\subsection{Susy analysis}
\medskip

A bosonic configuration of fields is supersymmetric iff the supersymmetry
variation of the fermionic fields, evaluated on the configuration, vanishes.
In our case the fermionic fields are two pseudo Majorana spinors
$\psi_{\lambda}$ and $\chi_i$ and their supersymmetry transformations
are given in appendix A (see eq. (\ref{apaseis})). In the configurations
considered in this section we are not exciting any coset scalar and,
therefore, we must take $P_{\mu ij}=0$ and $T_{ij}=\delta_{ij}$. Moreover,
through this paper we shall use the following
representation of the Dirac matrices:
\beq
\Gamma^{ \mu}\,=\,\gamma^{\mu}\,\,\otimes\,\II\,\,,
\,\,\,\,\,\,\,\,\,\,\,\,\,\,\,\,\,\,\,\,\,\,
\hat\Gamma^{ i}\,=\,\gamma_9\,\,\otimes\,\sigma^i\,\,,
\label{ocho}
\eeq
where $\gamma^{\mu}$ are eight dimensional Dirac matrices,
$\sigma^i$ are Pauli matrices and  $\gamma_9\,=\,i \gamma^{ 0}\,
\gamma^{ 1}\,\cdots\,\gamma^{7}$ ($\gamma_9^2\,=\,1$).
Actually, in what follows we shall denote by
$\{\Gamma_1, \Gamma_2,\Gamma_3\}$ the Dirac matrices along
the sphere $S^3$, by $\{\hat\Gamma_1, \hat\Gamma_2,\hat\Gamma_3\}$
the corresponding matrices along the
$SU(2)$ group manifold, whereas $\Gamma_7\equiv\Gamma_r$ will
correspond to the
$\Gamma$-matrix along the radial direction.

The first-order BPS equations we are trying to find are obtained by requiring
that $\delta\chi_i=\delta\psi_{\lambda}=0$ for some Killing spinor $\epsilon$,
which must satisfy some projection conditions. First of all, we shall impose
that:
\beq
\Gamma_{12}\,\epsilon\,=\,-\hat\Gamma_{12}\,\epsilon\,\,,
\,\,\,\,\,\,\,\,\,\,\,\,\,\,\,
\Gamma_{23}\,\epsilon\,=\,-\hat\Gamma_{23}\,\epsilon\,\,,
\,\,\,\,\,\,\,\,\,\,\,\,\,\,\,
\Gamma_{13}\,\epsilon\,=\,-\hat\Gamma_{13}\,\epsilon\,\,.
\label{nueve}
\eeq
Notice that in eq. (\ref{nueve}) the projections along the sphere $S^3$ and
the SU(2) group manifold are related. Actually, only two of these equations
are independent and, for example, the last one can be obtained from the
first two. Moreover, it follows from (\ref{nueve}) that:
\beq
\Gamma_1\hat\Gamma_1\epsilon\,=\,
\Gamma_2\hat\Gamma_2\epsilon\,=\,
\Gamma_3\hat\Gamma_3\epsilon\,\,.
\label{diez}
\eeq
These projections are imposed by the ambient Calabi--Yau three--fold in
which the three--cycle lives, from the conditions $J_{ab}~\epsilon =
\Gamma_{ab}~\epsilon$, where $J$ is the K\"ahler form. By using eqs.
(\ref{nueve})  and (\ref{diez}) to evaluate the right-hand side of
(\ref{apaseis}), together with the ansatz for the metric, dilaton and
gauge field, one gets some equations which give the
radial derivative of $\phi$, $h$ and
$\epsilon$. Actually, one arrives at the following equation for the radial
derivative of the dilaton:
\beq
\phi'\epsilon\,=\,{3\over 8}\,\Big[\,4\,(\,1-4g^2\,)\,e^{\phi-2h}\,-\,
e^{-\phi}\,\Big]\,\Gamma_r\,\hat\Gamma_{123}\,\epsilon\,+\,
3\,e^{\phi-h}\,g'\,\Gamma_1\,\hat\Gamma_1\,\epsilon\,\,,
\label{once}
\eeq
while the derivative of the function $h$ is:
\bear
h'\epsilon&=&2ge^{-h}\,
\Gamma_1\,\hat\Gamma_1\,\Gamma_r\,\hat\Gamma_{123}\,\epsilon\,-\,
{1\over 8}\,\Big[\,12\,(\,1-4g^2\,)\,e^{\phi-2h}\,+\,
e^{-\phi}\,\Big]\,\Gamma_r\,\hat\Gamma_{123}\,\epsilon\,-\,\rc\rc
&&-\,e^{\phi-h}\,g'\,\Gamma_1\,\hat\Gamma_1\,\epsilon\,\,.
\label{doce}
\eear
Moreover, the radial dependence of the spinor $\epsilon$ is determined by:
\bear
&&\partial_{r}\,\epsilon\,-\,{1\over 16}\,\Big[\,
4(\,1-4g^2\,)\,e^{\phi-2h}\,\,-\,e^{-\phi}\,\Big]\,
\Gamma_r\,\hat\Gamma_{123}\,\epsilon\,+\,
{5\over 2}\,e^{\phi-h}\,g'\,\Gamma_1\,\hat\Gamma_1\,\epsilon\,=\,0\,\,.
\label{trece}
\eear
In order to proceed further, we need to impose some additional condition  to
the spinor $\epsilon$. It is clear from the right-hand side of eqs.
(\ref{once})-(\ref{trece}) that we must specify the action on $\epsilon$ of
the radial projector $\Gamma_r\,\hat\Gamma_{123}$. The choice made in ref.
\cite{en} was to take $g=0$ and impose the condition
$\Gamma_r\,\hat\Gamma_{123}\,\epsilon\,=\,-\epsilon$. It is immediate to
verify that in this case our eqs. (\ref{once})--(\ref{trece}) reduce to
those obtained in ref. \cite{en}. Here we will not take any {\sl a priori}
particular value of
$\Gamma_r\,\hat\Gamma_{123}\,\epsilon$. Instead we will try to determine it
in general from eqs. (\ref{once})--(\ref{trece}). Notice that in our approach
$g$ will not be constant and, therefore, we will have to find a differential
equation which determines it.  It is clear  from  eq. (\ref{once}) that our
spinor $\epsilon$ must satisfy a relation of the sort:
\beq
\Gamma_r\,\hat\Gamma_{123}\,\epsilon\,=\,-
(\beta\,+\,\tilde\beta\,\Gamma_1\,\hat\Gamma_1\,)\,\epsilon\,\,,
\label{catorce}
\eeq
where $\beta$ and $\tilde\beta$ are functions of the radial coordinate $r$,
that can be easily extracted from eq. (\ref{once}), namely:
\bear
\beta&=&-{8\over 3}\,\,
{\phi'\over 4(1-4g^2)\,e^{\phi-2h}\,-\,e^{-\phi}}\,\,,\rc\rc
\tilde\beta&=&8\,\,
{e^{\phi-h}\,g'\over 4(1-4g^2)\,e^{\phi-2h}\,-\,e^{-\phi}}\,\,.
\label{quince}
\eear
Eq. (\ref{catorce}) is the kind of radial projection we are looking for.
We can get a consistency condition for this projection by noticing  that
$(\Gamma_r\,\hat\Gamma_{123})^2\epsilon\,=\,\epsilon$. Using the fact that 
$\{\Gamma_r\,\hat\Gamma_{123},\Gamma_1\,\hat\Gamma_1\}=0$, this condition is
simply :
\beq
\beta^2\,+\,\tilde\beta^2\,=\,1\,\,.
\label{dseis}
\eeq
By using in eq. (\ref{dseis}) the explicit values of $\beta$ and
$\tilde\beta$ given in eq. (\ref{quince}), one gets:
\beq
{(\phi')^2\over 9}\,+\,e^{2\phi-2h}\,(g')^2\,=\,
{1\over 64}\,\Big[\,4(1-4g^2)\,e^{\phi-2h}\,-\,e^{-\phi}\,\Big]^2\,\,,
\label{dsiete}
\eeq
which relates the radial derivatives of $\phi$ and $g$. Let us now
consider the equation for $h'$ written in eq. (\ref{doce}). By using the value
of  $\Gamma_r\,\hat\Gamma_{123}\,\epsilon$ given in eq. (\ref{catorce}),
and separating the terms with and without $\Gamma_1\,\hat\Gamma_1\,\epsilon$,
we get two equations:
\bear
&&h'\,=\,2ge^{-h}\,\tilde\beta\,+\,{1\over 8}\,
\Big[\,\,12(1-4g^2)\,e^{\phi-2h}\,+\,e^{-\phi}\,\Big]\,\beta\,\,,\rc\rc
&&2g\,e^{-h}\,\beta\,-\,{1\over 8}\,
\Big[\,\,12(1-4g^2)\,e^{\phi-2h}\,+\,e^{-\phi}\,\Big]\,\tilde\beta\,+\,
e^{\phi-h}\,g'\,=\,0\,\,.
\label{docho}
\eear
Moreover,  by using in the latter
the values of $\beta$ and $\tilde\beta$ given in
eq.  (\ref{quince}), we get the following relation between $g'$ and
$\phi'$:
\beq
g'\,=\,-{8g\over 3}\,
{e^{2h}\,\phi'\over 4(1-4g^2)\,e^{2\phi}\,+\,e^{2h}}\,\,.
\label{dnueve}
\eeq
Plugging back this equation in the consistency condition (\ref{dsiete}),
we can determine $\phi'$, $g'$, $\beta$ and $\tilde\beta$ in terms of
$\phi$, $g$ and $h$. Moreover, by substituting these results on the
first equation in (\ref{docho}), we get a first-order equation for $h$.
In order to write these
equations, let us define the function:
\beq
K\equiv
\sqrt{\Big(\,4\,(1-2g)^2\,e^{2\phi}+\,e^{2h}\,\Big)\,
\Big(\,4\,(1+2g)^2\,e^{2\phi}+\,e^{2h}\,\Big)}\,\,.
\label{veinte}
\eeq
Then, the BPS equations are:
\bear
\phi'&=&\,{3\over 8}\,\,{e^{-2h-\phi}\over K}\,\,\Big[\,
e^{4h}\,-\,16\,(1-4g^2)^2\,e^{4\phi}\,\Big]\,\,,\rc\rc
h'&=&{e^{-2h-\phi}\over 8K}\,\,\Big[\,
e^{4h}\,+\,16\,(1+4g^2)\,e^{2h+2\phi}\,+\,48\,
(1-4g^2)^2\,e^{4\phi}\,\Big]\,\,,\rc\rc
g'&=&{ge^{-\phi}\over K}\,\,\Big[\,4\,(1-4g^2)\,e^{2\phi}\,-\,
e^{2h}\,\Big]\,\,.
\label{vuno}
\eear
Notice that $g'=g=0$ certainly solves the last of these equations and, in this
case, the first two equations in (\ref{vuno}) reduce to the ones
written in ref. \cite{en}. Moreover, the system (\ref{vuno}) is identical to
that found in ref. \cite{CGLP1} by means of the superpotential method (see
appendix B). The solutions of (\ref{vuno}) have been obtained in ref.
\cite{CGLP1}, and they depend on two parameters (see below).

In order to have a clear interpretation   of the radial projection we are
using, let us notice that, due to the constraint (\ref{dseis}), we can
represent
$\beta$ and $\tilde\beta$ as:
\beq
\beta\,=\,\cos\alpha\,\,,
\,\,\,\,\,\,\,\,\,\,\,\,\,\,
\tilde\beta\,=\,\sin\alpha\,\,.
\label{vdos}
\eeq
Moreover, by substituting the value of $\phi'$ and $g'$ given by the
first-order equations (\ref{vuno}) into the definition of $\beta$ and
$\tilde\beta$ (eq. (\ref{quince})), one arrives at:
\beq
\tan\alpha\,=\,8g\,{e^{\phi+h}\over 4(1-4g^2)\,e^{2\phi}\,+\,e^{2h}}\,\,.
\label{vtres}
\eeq
Then, by using the representation (\ref{vdos}), it is immediate to rewrite
eq. (\ref{catorce}) as:
\beq
\Gamma_r\,\hat\Gamma_{123}\,\epsilon\,=\,-
e^{\alpha\Gamma_1\hat\Gamma_1}\,\,\epsilon\,\,.
\label{vcuatro}
\eeq
Since $\{\Gamma_r\,\hat\Gamma_{123},\Gamma_1\,\hat\Gamma_1\}=0$,
eq.  (\ref{vcuatro}) can be solved as:
\beq
\epsilon\,=\,e^{-{1\over 2}\alpha\Gamma_1\hat\Gamma_1}
\,\,\epsilon_0\,\,,
\label{vcinco}
\eeq
where $\epsilon_0$ is a spinor satisfying the standard radial projection
condition with $\alpha=0$, \ie:
\beq
\Gamma_r\,\hat\Gamma_{123}\,\epsilon_0\,=\,-\epsilon_0\,\,.
\label{vseis}
\eeq
To determine completely $\epsilon_0$ we must use eq. (\ref{trece}), which
dictates the radial dependence of the Killing spinor. Actually, by using the
first-order equations (\ref{vuno}) one can compute $\partial_r\,\alpha$ from
eq. (\ref{vtres}). The result is remarkably simple, namely:
\beq
\partial_r\,\alpha\,=\,6\,e^{\phi-h}\,g'\,\,.
\label{vsiete}
\eeq
By using eqs. (\ref{vcuatro}) and (\ref{vsiete}) in eq. (\ref{trece}), one
can verify that $\epsilon_0$ satisfies the equation:
\beq
\partial_r\,\epsilon_0\,=\,{\phi'\over 6}\,\epsilon_0\,\,,
\label{vocho}
\eeq
which can be immediately integrated as:
\beq
\epsilon_0\,=\,e^{{\phi\over 6}}\,\eta\,\,,
\label{vnueve}
\eeq
with $\eta$ being a constant spinor. Thus, after collecting all
results, it follows that $\epsilon$ can be written as:
\beq
\epsilon\,=\,e^{{\phi\over 6}}\,
e^{-{1\over 2}\alpha\Gamma_1\hat\Gamma_1}\,\eta\,\,.
\label{treinta}
\eeq
The projections conditions satisfied by  $\eta$ are simply:
\beq
\Gamma_{12}\,\hat\Gamma_{12}\,\eta\,=\,\eta\,\,,
\,\,\,\,\,\,\,\,\,\,\,\,\,\,\,\,\,\,\,\,\,\,\,\,
\Gamma_{23}\,\hat\Gamma_{23}\,\eta\,=\,\eta\,\,,
\,\,\,\,\,\,\,\,\,\,\,\,\,\,\,\,\,\,\,\,\,\,\,\,
\Gamma_{r}\,\hat\Gamma_{123}\,\eta\,=\,-\eta\,\,.
\label{tuno}
\eeq

In order to find out the meaning  of the  phase $\alpha$, let us notice
that, by using the representation (\ref{ocho}) for the
$\Gamma$-matrices, one easily proves that:
\beq
\Gamma_{x^0\cdots x^3}\,\Gamma_{123}\,\Gamma_r\,\hat\Gamma_{123}\,=\,-1\,\,.
\label{tdos}
\eeq
From eqs. (\ref{nueve}) and (\ref{tdos}), it is straightforward to verify
that the radial projection (\ref{vcuatro}) can be written as:
\beq
\Gamma_{x^0\cdots x^3}\,\big(\,\cos\alpha\Gamma_{123}\,-\,
\sin\alpha\hat\Gamma_{123}\,\big)\,\epsilon\,=\,\epsilon\,\,,
\label{ttres}
\eeq
which is the projection corresponding to a D6--brane wrapped on a
three--cycle,
which is non-trivially embedded in the two three--spheres, with $\alpha$
measuring the contribution of each sphere. This equation must be understood
as seen from the uplifted perspective. The case $\alpha = 0$ corresponds
to the D6--brane wrapping a three--sphere that is fully contained in the
eight--dimensional spacetime where supergravity lives, and has been
studied earlier \cite{en}. Notice that $\alpha = \pi/2$ is not a solution
of the system. This is an important consistency check as this would mean
that the D6--brane is not wrapping a three--cycle contained in the
eight--dimensional spacetime and the twisting would make no sense.
However, solutions that asymptotically approach $\alpha = \pi/2$ are
possible in principle. In the next subsection we will describe a
quantity for which the rotation by the angle $\alpha$ plays an
important role.

\subsection{The calibrating three-form}
\medskip
Given a solution of the BPS equations (\ref{vuno}), one can get an eleven
dimensional metric $ds^2_{11}$ by means of the uplifting formula
(\ref{apacuatro}). The condition (\ref{cinco}) ensures that the
corresponding eleven dimensional manifold is a direct product of four
dimensional Minkowski space and a seven dimensional manifold, \ie:
\beq
ds^2_{11}\,=\,dx^2_{1,3}\,+\,ds^2_7\,=\,
dx^2_{1,3}\,+\,\sum_{A=1}^{7}\,\big(\,e^{A}\,)^2\,\,,
\label{tcuatro}
\eeq
where we have written $ds^2_7$ in terms of  a basis of one-forms
$e^{A}$ ($A=1,\cdots, 7)$. It follows from (\ref{apacuatro}) that this basis
can be taken as:
\bear
e^{i}&=&{1\over 2}\,e^{h-{\phi\over 3}}\,\,w^{i}\,,
\,\,\,\,\,\,\,\,\,\,(i=1,2,3)\,\,,\rc\rc
e^{3+i}&=&2\,e^{{2\phi\over 3}}\,
\big(\,\tilde w^i\,+\,(\,g-{1\over 2}\,)\,w^{i}\,\,\big)\,,
\,\,\,\,\,\,\,\,\,\,(i=1,2,3)\,\,,\rc\rc
e^{7}&=&e^{-{\phi\over 3}}\,dr\,\,.
\label{tcinco}
\eear
It is a well-known fact that a manifold of $G_2$ holonomy is endowed with a
calibrating three-form $\Phi$, which must be closed and co-closed  with
respect
to the seven dimensional metric $ds^2_7$. We shall denote by
$\phi_{ABC}$  the components of $\Phi$ in the basis $(\ref{tcinco})$, namely:
\beq
\Phi\,=\,{1\over 3!}\,\phi_{ABC}\,e^{A}\wedge e^{B}\wedge e^{C}\,\,.
\label{tseis}
\eeq
The relation between $\Phi$ and the Killing spinors of the metric is also
well-known. Indeed, let $\tilde \epsilon$ be the Killing spinor uplifted to
eleven dimensions, which in terms of $\epsilon$ is simply
$\tilde \epsilon= e^{-{\phi\over 6}}\,\epsilon$. Then, one has:
\beq
\phi_{ABC}\,=\,i\,\tilde \epsilon^{\dagger}\,\Gamma_{ABC}\,\tilde
\epsilon\,\,.
\label{tsiete}
\eeq
By using the relation between $\epsilon$ and the constant spinor
$\eta$, one can rewrite eq. (\ref{tsiete}) as:
\beq
\phi_{ABC}\,=\,i\, \eta^{\dagger}\,
e^{{1\over 2}\alpha\Gamma_1\hat\Gamma_1}
\Gamma_{ABC}\, e^{-{1\over 2}\alpha\Gamma_1\hat\Gamma_1}\eta\,\,.
\label{tocho}
\eeq
Let us now denote by $\phi_{ABC}^{(0)}$  the  above matrix element when
$\alpha=0$, \ie:
\beq
\phi_{ABC}^{(0)}\,=\,i\, \eta^{\dagger}\,\Gamma_{ABC}\,\eta\,\,.
\label{tnueve}
\eeq
It is not difficult to obtain the non-zero matrix elements (\ref{tnueve}).
Recall that $\eta$ is characterized as an eigenvector of the set of projection
operators written in eq. (\ref{tuno}). Thus, if ${\cal O}$ is an operator
which anticommutes with these projectors, ${\cal O}\eta$ and $\eta$ are
eigenvectors of the projectors with different eigenvalues and, therefore,
they are orthogonal (\ie\ $\eta^{\dagger}{\cal O}\eta=0$). Moreover, by
using the projection conditions (\ref{tuno}), one can relate the
non-vanishing matrix elements to $\eta^{\dagger}\,\Gamma_{123}\,\eta$.
If we normalize $\eta$ such that $i\, \eta^{\dagger}\,\Gamma_{123}\,
\eta\,=\,1$ and if $\hat i\,=\,i+3$ for $i=1,2,3$, one can easily prove
that the non-zero $\phi_{ijk}^{(0)}$'s are:
\beq
\phi_{ijk}^{(0)}\,=\,\epsilon_{ijk}\,\,,
\,\,\,\,\,\,\,\,\,\,\,\,\,\,\,\,\,\,
\phi_{i\hat j\hat k}^{(0)}\,=\,-\epsilon_{ijk}\,\,,
\,\,\,\,\,\,\,\,\,\,\,\,\,\,\,\,\,\,
\phi_{7i\hat j}^{(0)}\,=\,\delta_{ij}\,\,,
\label{cuarenta}
\eeq
as in \cite{ads}. By expanding the exponential in (\ref{tocho}) and
using (\ref{cuarenta}), it is
straightforward to find the different components of $\Phi$ for arbitrary
$\alpha$. Actually, one can write the result as:
\bear
\Phi&=&e^{7}\,\wedge\,\big(\,e^{1}\,\wedge\,e^{4}\,+\,
e^{2}\,\wedge\,e^{5}\,+\,e^{3}\,\wedge\,e^{6}\,\big)\,+\,\rc\rc
&&+\,\big(\,e^{1}\cos\alpha+e^{4}\sin\alpha\,\big)\,\wedge\,
\big(\,e^{2}\wedge e^{3}-e^{5}\wedge e^{6}\,\big)\,+\rc\rc
&&+\,\big(\,-e^{1}\sin\alpha+e^{4}\cos\alpha\,\big)\,\wedge\,
\big(\,e^{3}\wedge e^{5}-e^{2}\wedge e^{6}\,\big)\,\,,
\label{cuno}
\eear
which shows that the effect on $\Phi$ of introducing the phase $\alpha$ is
just a (radial dependent) rotation in the $(e^{1}, e^{4})$ plane
(alternatively, the same
expression can be written as a rotation in the $(e^{2}, e^{5})$ or
$(e^{3}, e^{6})$ plane). As mentioned above, $\Phi$
should be closed and co-closed:
\beq
d\Phi=0\,\,,
\,\,\,\,\,\,\,\,\,\,\,\,\,\,\,\,\,\,
d*_7\Phi=0\,\,,
\label{cdos}
\eeq
where $*_7$ denotes the Hodge dual in the seven dimensional metric. There
is an immediate consequence of this fact which we shall now exploit. Let us
denote by $p$ and $q$ the components of $\Phi$ along the volume forms of
the two three spheres, \ie:
\beq
\Phi\,=\,p\,w^1\wedge w^2\wedge w^3\,+
\,q\,\tilde w^1\wedge \tilde w^2\wedge \tilde w^3\,+\,\cdots\,\,.
\label{ctres}
\eeq
From the condition $d\Phi=0$, it follows immediately that $p$ and $q$ must be
constants of motion. By plugging the explicit expression of the forms $e^A$,
given in eq. (\ref{tcinco}), on the right-hand side of eq. (\ref{cuno}), one
easily gets $p$ and $q$ in terms of $\phi$, $h$ and $g$. The result is:
\bear
p&=&{1\over 8}\,\Big[\,e^{3h-\phi}\,-\,12\,e^{h+\phi}\,
\,(\,1-2g\,)^2\,\Big]\,\cos\alpha\,-\,
{1\over 4}\,(1-2g)\,
\Big[\,3e^{2h}\,-\,4\,e^{2\phi}\,(\,1-2g\,)^2\,\Big]
\,\sin\alpha\,\,,\rc\rc
q&=&-8e^{2\phi}\,\sin\alpha\,\,.
\label{ccuatro}
\eear
Notice that $\alpha = 0$ implies $q = 0$ which is precisely the case
studied in \cite{en}.
By explicit calculation one can check that $p$ and $q$ are constants as a
consequence of the BPS equations. Actually, by using (\ref{vuno})
one can show that, indeed, $\Phi$ is closed and co-closed as it should.

To finish this section, let us write down  the general solution of the
first-order system (\ref{vuno}), which was obtained in ref. \cite{CGLP1}. This
solution is expressed in terms of a new radial variable $\rho$ and of the two
following functions $Y(\rho)$ and $F(\rho)$:
\bear
Y(\rho)&\equiv&\rho^2\,-\,2(2p+q)\rho\,+\,4p(p+q)\,\,,\rc\rc
F(\rho)&\equiv&3\rho^4\,-\,8(2p+q)\rho^3\,+\,24p(p+q)\,\rho^2\,-\,
16p^2(p+q)^2\,\,.
\label{ccinco}
\eear
Notice that $Y(\rho)$ and $F(\rho)$ also depend on the two constants
$p$ and $q$. The seven dimensional metric takes the form:
\beq
ds^2_7\,=\,F^{-{1\over 3}}\,d\rho^2\,+\,
{1\over 4}\,F^{{2\over 3}}\,Y^{-1}\,(\,w^i\,)^2\,+\,
F^{-{1\over 3}}\,Y\,
\Bigg(\,\tilde w^i\,-\,
\big(\,{1\over 2}\,+\,{q\rho\over  Y}\big)\,w^i\,\Bigg)^2\,\,.
\label{cseis}
\eeq
The analysis of the metrics (\ref{cseis}) has been carried out in ref.
\cite{CGLP1}. It turns out that only in three cases ($p=0$, $q=0$ and
$p=-q$) the metric  (\ref{cseis}) is non-singular. The first two cases
are related by the so-called flop transformation, which is a $\ZZ_2$
action that exchanges $w^i$ and $\tilde w^i$, while the $p=-q$ case
is flop invariant. It is interesting to point out that, as $g \to 0$
when $\rho \to \infty$, the gauge field (\ref{seis}) asymptotically
approaches that used in \cite{en} to perform the twisting. This is
in line with the fact that the twisting just fixes the value of the
gauge field where the gauge theory lives, {\em i.e.}, at infinity.

\setcounter{equation}{0}
\section{D6-branes on a squashed 8d metric}
\medskip
In this section we are going to generalize the analysis performed in
section 2 to a much more general situation, in which the eight dimensional
metric takes the form:
\beq
ds^2_8\,=\,e^{{2\phi\over 3}}\,dx_{1,3}^2\,+\,{1\over 4}\,
e^{2h_i}\,(\,w^i\,)^2\,+\,dr^2\,\,.
\label{csiete}
\eeq
Notice that in the ansatz (\ref{csiete}) we have already implemented the
condition (\ref{cinco}), which ensures that we are going to have a direct
product of four dimensional Minkowski space and a seven dimensional manifold
in the uplift to eleven dimensions. As in the previous case, we are going
to switch on a SU(2) gauge field potential with components along the
squashed $S^3$. The ansatz we shall adopt for this potential is:
\beq
A^i\,=\,G_i\,w^i\,\,,
\label{cocho}
\eeq
which depends on three functions $G_1$,  $G_2$ and  $G_3$. It should be
understood that there is no sum on the right-hand side of eq.
(\ref{cocho}). Moreover, we shall excite coset scalars in the diagonal and,
therefore, the corresponding $L_{\alpha}^i$ matrix will be taken as:
\beq
L_{\alpha}^i\,=\,{\rm diag}\,
(\,e^{\lambda_1}\,,\,e^{\lambda_2}\,,\,e^{\lambda_3}\,)\,\,,
\,\,\,\,\,\,\,\,\,\,\,\,
\lambda_1\,+\,\lambda_2\,+\,\lambda_3\,=\,0\,\,.
\label{cnueve}
\eeq
The matrices $P_{\mu ij}$ and $Q_{\mu ij}$ defined in appendix A (eq.
(\ref{apauno}) ) are easily evaluated from eqs. (\ref{cocho}) and
(\ref{cnueve}). Written as differential forms, they are:
\beq
P_{ij}\,+\,Q_{ij}\,=\,
\pmatrix{d\lambda_1&&-A^{3}\,e^{\lambda_{21}}&&A^{2}\,e^{\lambda_{31}}\cr\cr
         A^{3}\,e^{\lambda_{12}}&&d\lambda_2&&-A^{1}\,e^{\lambda_{32}}\cr\cr
        -A^{2}\,e^{\lambda_{13}}&&A^{1}\,e^{\lambda_{23}}&&d\lambda_3}\,\,,
\label{cincuenta}
\eeq
where $\lambda_{ij}\,\equiv\,\lambda_i\,-\,\lambda_j$ and
$P_{ij}$ ($Q_{ij}$) is the symmetric (antisymmetric) part of the matrix
appearing on the right-hand side of (\ref{cincuenta}). Notice that our present
ansatz depends on nine functions, since there are only two independent
$\lambda_i$'s (see eq. (\ref{cnueve})). On the other hand, it would be
convenient in what follows to define the following combinations of these
functions:
\bear
M_1\,\equiv &\,e^{\phi+\lambda_1-h_2-h_3}\,(\,G_1\,+\,G_2\,G_3)\,\,,\rc\rc
M_2\,\equiv&\,e^{\phi+\lambda_2-h_1-h_3}\,(\,G_2\,+\,G_1\,G_3)\,\,,\rc\rc
M_3\,\equiv&\,e^{\phi+\lambda_3-h_1-h_2}\,(\,G_3\,+\,G_1\,G_2)\,\,.
\label{ciuno}
\eear

\subsection{Susy analysis}
\medskip

With the setup just described, and the experience acquired in the previous
section, we will now attack the problem of finding supersymmetric
configurations
for this more general ansatz. As before, we must guarantee that
$\delta\chi_i=\delta\psi_{\lambda}=0$ for some spinor $\epsilon$. We begin by
imposing again the angular projection condition (\ref{nueve}). Then, the
equation $\delta\chi_1=0$ yields:
\bear
\Big(\,{1\over 2}\lambda_1'+{1\over 3}\phi'\,\Big)\,\epsilon\,&=&\,
e^{\phi+\lambda_1-h_1}\,G_1'\,\Gamma_1\hat\Gamma_1\,\epsilon\,-\,
2\Big[\,M_1\,-\,{1\over 16}\,e^{-\phi}\,
(\,e^{2\lambda_1}-e^{2\lambda_2}-e^{2\lambda_3}\,)\,\Big]\,
\Gamma_r\hat\Gamma_{123}\,\epsilon\,-\,\rc\rc
&&-\Big[\,e^{-h_2}\,G_2\,\sinh\lambda_{13}\,+\,
e^{-h_3}\,G_3\,\sinh\lambda_{12}\,\Big]\,
\Gamma_1\hat\Gamma_1\,\Gamma_r\hat\Gamma_{123}\,\epsilon\,\,,
\label{cidos}
\eear
and, obviously, $\delta\chi_2=\delta\chi_3=0$ give rise to
other two similar equations which are obtained by permutation of the indices
$(1,2,3)$ in eq. (\ref{cidos}). Adding these three equations and using eq.
(\ref{diez}) and the fact that  $\lambda_1+\lambda_2+\lambda_3=0$, we get the
following equation for $\phi'$:
\bear
\phi'\,\epsilon\,&=&\,e^{\phi}\,\Big[\,
e^{\lambda_1-h_1}\,G_1'\,+\,e^{\lambda_2-h_2}\,G_2'\,+\,
e^{\lambda_3-h_3}\,G_3'\,\Big]\Gamma_1\hat\Gamma_1\,\epsilon\,-\,\rc\rc
&&-2\Big[\,M_1+M_2+M_3+{1\over 16}\,e^{-\phi}\,
\big(\,e^{2\lambda_1}\,+\,e^{2\lambda_2}\,+\,e^{2\lambda_3}\,\big)\,\Big]\,
\Gamma_r\hat\Gamma_{123}\,\epsilon\,\,.
\label{citres}
\eear
It can be checked that this same equation is obtained from the variation of
the gravitino components along the unwrapped directions. Moreover, it
follows from  eq. (\ref{citres}) that
$\Gamma_r\,\hat\Gamma_{123}\,\epsilon$ has
the same structure as in eq. (\ref{catorce}), where now $\beta$ and
$\tilde\beta$ are given by:
\bear
\beta&=&\,{8\,\phi'\over
16\,(\,M_1+M_2+M_3)\,+\,e^{-\phi}\,
(\,e^{2\lambda_1}\,+\,e^{2\lambda_2}\,+\,e^{2\lambda_3}\,)}\,\,,\rc\rc\rc
\tilde\beta&=&\,-\,{8\,e^{\phi}\,(\,
e^{\lambda_1-h_1}\,G_1'\,+\,e^{\lambda_2-h_2}\,G_2'\,+\,
e^{\lambda_3-h_3}\,G_3'\,)
\over
16\,(\,M_1+M_2+M_3)\,+\,e^{-\phi}\,
(\,e^{2\lambda_1}\,+\,e^{2\lambda_2}\,+\,e^{2\lambda_3}\,)}\,\,.
\label{cicuatro}
\eear
It is also immediate to see that in the present case $\beta$ and $\tilde\beta$
must also satisfy the constraint (\ref{dseis}). Thus, in this case we are
going to have the same type of radial projection as in the round metric
of section 2.
Actually, we shall obtain a set of first-order differential equations in
terms of $\beta$ and $\tilde\beta$ and then we shall find some consistency
conditions which, in particular, allow to determine  the  values of
$\beta$ and $\tilde\beta$. From this point of view it is straightforward to
write the equation for $\phi'$. Indeed, from the definition of $\beta$
(eq. (\ref{cicuatro})), one has:
\beq
\phi'\,=\,\Big[\,2\,(\,M_1\,+\,M_2\,+\,M_3)\,+\,{1\over 8}\,
e^{-\phi}\,(\,e^{2\lambda_1}\,+\,e^{2\lambda_2}\,+\,e^{2\lambda_3}\,)\,
\Big]\,\beta\,\,.
\label{cicinco}
\eeq
In order to obtain the equation for $\lambda_i'$ and $G_i'$, let us consider
again the equation derived from the condition $\delta\chi_i=0$
(eq. (\ref{cidos})). Plugging the projection condition on the right-hand side
of eq. (\ref{cidos}), using the value of $\phi'$ displayed in eq.
(\ref{cicinco}), and considering the terms without $\Gamma_1\,\hat\Gamma_1$,
one gets the equation for $\lambda_1'$, namely:
\bear
\lambda_1'&=&{4\over 3}\,
\Big[\,2M_1\,-\,M_2\,-\,M_3\,-\,{1\over 8}\,e^{-\phi}\,\big(\,
2e^{2\lambda_1}\,-\,e^{2\lambda_2}\,-\,e^{2\lambda_3}\,\big)\,
\Big]\,\beta\,-\,\rc\rc
&&-\,2\,\Big[\,e^{-h_2}\,G_2\,\sinh\lambda_{13}\,+\,
e^{-h_3}\,G_3\,\sinh\lambda_{12}\,\Big]\,\tilde\beta\,\,.
\label{ciseis}
\eear
while the terms with $\Gamma_1\,\hat\Gamma_1$ of eq. (\ref{cidos}) yield the
equation for $G_1'$:
\bear
e^{\phi\,+\,\lambda_1-h_1}\,G_1'&=&\Big[\,
-2M_1\,+\,{1\over 8}\,e^{-\phi}\,
(\,e^{2\lambda_1}\,-\,e^{2\lambda_2}\,-\,e^{2\lambda_3}\,)
\,\Big]\,\tilde\beta\,-\,\rc\rc
&&-\,\Big[\,
e^{-h_2}\,G_2\,\sinh\lambda_{13}\,+\,
e^{-h_3}\,G_3\,\sinh\lambda_{12}\,\Big]\,\beta\,\,.
\label{cisiete}
\eear
By cyclic permutation of eqs. (\ref{ciseis}) and (\ref{cisiete}) one obtains
the first-order differential equations of  $\lambda_i'$ and $G_i'$ for
$i=2,3$.

It remains to obtain the equation for  $h_i'$. With this purpose let us
consider the supersymmetric variation of the gravitino components along the
sphere. One gets:
\bear
h_1'\,\epsilon&=&-{1\over 3}\,e^{\phi}\,
\Big[\,5\,e^{\lambda_1-h_1}\,G_1'\,-\,e^{\lambda_2-h_2}\,G_2'\,-\,
e^{\lambda_3-h_3}\,G_3'\,\Big]\,\Gamma_1\,\hat\Gamma_1\,\epsilon
\,-\,\rc\rc
&&-{1\over 3}\,\Big[\,2\,(M_1-5M_2-5M_3)\,+\,{1\over 8}\,
e^{-\phi}\,(\,e^{2\lambda_1}\,+\,e^{2\lambda_2}\,+\,e^{2\lambda_3}\,)
\,\Big]\,\Gamma_r\,\hat\Gamma_{123}\,\epsilon\,-\,\rc\rc
&&-\,\Big[\,{e^{2h_1}\,-\,e^{2h_2}\,-\,e^{2h_3}\over
e^{h_1+h_2+h_3}}\,-\,2e^{-h_1}\,G_1\,\cosh\lambda_{23}\,\Big]\,
\Gamma_1\,\hat\Gamma_1\,\Gamma_r\,\hat\Gamma_{123}\,\epsilon\,\,,
\label{ciocho}
\eear
and two other equations obtained by cyclic permutation. By considering
the terms without $\Gamma_1\,\hat\Gamma_1$ in eq. (\ref{ciocho})
we get the desired first-order equation for $h_1'$, namely:
\bear
h_1'&=&
{1\over 3}\,\Big[\,2\,(M_1-5M_2-5M_3)\,+\,{1\over 8}\,
e^{-\phi}\,(\,e^{2\lambda_1}\,+\,e^{2\lambda_2}\,+\,e^{2\lambda_3}\,)
\,\Big]\,\beta\,-\,\rc\rc
&&-\,\Big[\,{e^{2h_1}\,-\,e^{2h_2}\,-\,e^{2h_3}\over
e^{h_1+h_2+h_3}}\,
-\,2e^{-h_1}\,G_1\,\cosh\lambda_{23}\,\Big]\,
\tilde\beta\,\,.
\label{cinueve}
\eear
On the other hand, the terms with $\Gamma_1\,\hat\Gamma_1$ of
eq. (\ref{ciocho}) give rise to new equations for the $G_i'$'s:
\bear
&&e^{\phi}\,
\Big[\,5\,e^{\lambda_1-h_1}\,G_1'\,-\,e^{\lambda_2-h_2}\,G_2'\,-\,
e^{\lambda_3-h_3}\,G_3'\,\Big]\,=\,\rc\rc
&&=\,\Big[\,2\,(M_1-5M_2-5M_3)\,+\,{1\over 8}\,
e^{-\phi}\,(\,e^{2\lambda_1}\,+\,e^{2\lambda_2}\,+\,e^{2\lambda_3}\,)
\,\Big]\,\tilde\beta\,+\,\rc\rc
&&+\,3\,\Big[\,{e^{2h_1}\,-\,e^{2h_2}\,-\,e^{2h_3}\over
e^{h_1+h_2+h_3}}\,-\,2e^{-h_1}\,G_1\,\cosh\lambda_{23}\,\Big]\,\beta\,\,.
\label{sesenta}
\eear
This equation (and the other two obtained by cyclic permutation) must be
compatible with  the equation for $G_i'$ written in eq.
(\ref{cisiete}). Actually, by
substituting in eq. (\ref{sesenta}) the value of $G_i'$ given by eq.
(\ref{cisiete}), and by combining appropriately the equations so obtained,
we arrive at three algebraic relations of the type:
\beq
{\cal A}_i\,\beta\,-\,{\cal B}_i\,\tilde\beta\,=\,0\,\,,
\label{suno}
\eeq
where ${\cal A}_1$ and ${\cal B}_1$ are given by:
\bear
{\cal A}_1&=&e^{h_1-h_2-h_3}\,+\,
e^{\lambda_1-\lambda_3-h_2}\,G_2\,+\,
e^{\lambda_1-\lambda_2-h_3}\,G_3\,\,,\rc\rc
{\cal B}_1&=&-4M_1\,+\,{1\over 4}\,e^{-\phi+2\lambda_1}\,\,,
\label{sdos}
\eear
while the values of ${\cal A}_i$ and ${\cal B}_i$ for $i=2,3$ are obtained
from (\ref{sdos}) by cyclic permutation. Notice that the above relations
do not
involve derivatives of the fields and, in particular, they allow to obtain
the values of $\beta$ and  $\tilde\beta$. Indeed, by using the constraint
$\beta^2\,+\,\tilde\beta^2=1$,  and eq. (\ref{suno})  for $i=1$, we get:
\beq
\beta\,=\,{{\cal B}_1\over \sqrt{{\cal A}_1^2\,+\,{\cal B}_1^2}}\,\,,
\,\,\,\,\,\,\,\,\,\,\,
\tilde\beta\,=\,{{\cal A}_1\over \sqrt{{\cal A}_1^2\,+\,{\cal B}_1^2}}\,\,.
\label{stres}
\eeq
Moreover, it is clear from (\ref{suno}) that the ${\cal A}_i$'s and ${\cal
B}_i$'s must satisfy the following  consistency conditions:
\beq
{\cal A}_i\,{\cal B}_j\,=\,{\cal A}_j\,{\cal B}_i\,\,,
\,\,\,\,\,\,\,\,\,\,\,\,
(i\not= j)\,\,.
\label{scuatro}
\eeq
Eq. (\ref{scuatro}) gives two independent algebraic constraints that the
functions of our generic ansatz must satisfy if we demand it  to be a
supersymmetric
solution. Notice that these constraints are trivially satisfied in the round
case of section 2. On the other hand, if we adopt the radial projection of
refs. \cite{en, hs1}, \ie\ when  $\beta=1$ and $\tilde\beta=0$, they imply
that ${\cal A}_i=0$ (see eq. (\ref{suno})), this leading precisely to the
values of
the SU(2) gauge potential used in those references. Moreover, by using
eq. (\ref{suno}), the differential equation satisfied by the $G_i$'s
can be simplified. One gets:
\bear
e^{\phi\,+\,\lambda_1-h_1}\,G_1&=&{1\over 2}\,\,\Big[\,
e^{h_1-h_2-h_3}\,+\,e^{\lambda_3-\lambda_1-h_2}\,G_2\,+\,
e^{\lambda_2-\lambda_1-h_3}\,G_3\,\Big]\,\beta\,-\,\rc\rc 
&&\,-\,{e^{-\phi}\over 8}\,\big(\,e^{2\lambda_2}+e^{2\lambda_3}\,\big)\,
\tilde\beta\,\,,
\label{scinco}
\eear
and similar expressions for $G_2$ and $G_3$.

Let us now parametrize $\beta$ and $\tilde\beta$ as in eq. (\ref{vdos}),
\ie\ $\beta\,=\,\cos\alpha$, $\tilde\beta\,=\,\sin\alpha$. Then, it follows
from
eq. (\ref{stres}) that one has:
\beq
\tan\alpha\,=\,{{\cal A}_1\over {\cal B}_1}\,=\,
{{\cal A}_2\over {\cal B}_2}\,=\,
{{\cal A}_3\over {\cal B}_3}\,\,.
\label{sseis}
\eeq
Notice that by taking $\alpha = 0$, eq. (\ref{scinco}) precisely leads
to the expression for the gauge field in terms of scalar fields used in
\cite{hs1} to perform the twisting. Moreover, the radial projection
condition can be written as in eq. (\ref{vcuatro}) and, thus, the
natural solution to the Killing spinor equations is just the one written
in eq. (\ref{treinta}), where $\eta$ is a constant spinor satisfying the
conditions (\ref{tuno}). To check that this is the case, one can plug the
expression of $\epsilon$  given in eq. (\ref{treinta}) in the equation
arising from  the supersymmetric variation of the radial component of
the gravitino. It turns out that this equation is satisfied provided
$\alpha$ satisfies the equation:
\beq
\partial_r\alpha\,=\,-\,\Big[\,
4\,\big(\,M_1\,+\,M_2\,+\,M_3\,\big)\,+\,
{1\over 4}\,e^{-\phi}\,\big(\,
e^{2\lambda_1}\,+\,e^{2\lambda_2}\,+\,e^{2\lambda_3}\,\big)\,\Big]\,
\sin\alpha\,\,.
\label{ssiete}
\eeq
In general, this equation for $\alpha$ does not follow from the first-order
equations and the algebraic constraints we have found. Actually, by using the
value of $\alpha$ given in eq. (\ref{sseis}) and the first-order equations to
evaluate the left-hand side of eq. (\ref{ssiete}), we could derive a third
algebraic constraint. However, this new constraint is rather complicated.
Happily,  we will not need to do this explicitly since eq. (\ref{ssiete})
will serve to our purposes.

\subsection{The calibrating three-form}
\medskip

In order to find the calibrating three-form $\Phi$ in this case, let us
take the following vierbein basis:
\bear
e^{i}&=&{1\over 2}\,e^{h_i-{\phi\over 3}}\,\,w^{i}\,,
\,\,\,\,\,\,\,\,\,\,(i=1,2,3)\,\,,\rc\rc
e^{3+i}&=&2\,e^{{2\phi\over 3}+\lambda_i}\,
\big(\,\tilde w^i\,+\,G_i\,w^{i}\,\,\big)\,,
\,\,\,\,\,\,\,\,\,\,(i=1,2,3)\,\,,\rc\rc
e^{7}&=&e^{-{\phi\over 3}}\,dr\,\,,
\label{socho}
\eear
which is the natural one for the uplifted metric. The different components of
$\Phi$ can be computed by using eq. (\ref{tseis}) and it is obvious from the
form of the projection that the result is just the one given in eq.
(\ref{cuno}), where now $\alpha$ is given by eq. (\ref{sseis}) and the
one-forms $e^A$ are the ones written in eq. (\ref{socho}). If, as in eq.
(\ref{ctres}), $p$ and $q$ denote the components of
$\Phi$ along the two three spheres, it follows from the closure of $\Phi$ that
$p$ and $q$ should be constants of motion. By plugging the expressions of the
$e^A$'s, taken from eq. (\ref{socho}), on the right-hand side of eq.
(\ref{cuno}), one can find the explicit expressions of $p$ and $q$. The result
is:
\bear
p&=&{1\over 8}\,\Big[\,e^{h_1+h_2+h_3-\phi}\,-\,
16e^{\phi}\,\big(\,e^{h_1-\lambda_1}G_2G_3+\,
e^{h_2-\lambda_2}G_1G_3+\,e^{h_3-\lambda_3}G_1G_2\,\big)\Big]\,
\cos\alpha\,+\,\rc\rc
&&+\,{1\over 2}\,\Big[\,
e^{h_2+h_3+\lambda_1}G_1\,+\,e^{h_1+h_3+\lambda_2}G_2\,+\,
e^{h_1+h_2+\lambda_3}G_3\,-\,16e^{2\phi}\,G_1G_2G_3\,\Big]\,
\sin\alpha\,\,,\rc\rc
q&=&-8e^{2\phi}\sin\alpha \,\,.
\label{snueve}
\eear
It is a simple exercise to verify that, when restricted to the round case
studied above, the expressions of $p$ and $q$ given in eq.
(\ref{snueve}) coincide with those written in eq. (\ref{ccuatro}). Moreover,
the proof of the constancy of $p$ and $q$ can be performed by combining
appropriately the
first-order equations and the constraints. Actually, by using eq.
(\ref{cicinco}) to compute the radial derivative of $q$ in eq.
(\ref{snueve}), it
follows that the condition  $\partial_r q=0$ is equivalent to eq.
(\ref{ssiete}). Although the proof of $\partial_r p=0$ is much more involved,
one can demonstrate that $p$ is indeed constant by using the BPS equations and
the constraints (\ref{scuatro}) and (\ref{ssiete}).

\setcounter{equation}{0}
\section{The Hitchin system }
\medskip

A simple counting argument can be used to determine the  number of
independent functions left out by the constraints. Indeed, we have already
mentioned that our ansatz depends on nine functions. However, we have found
two constraints in eq. (\ref{scuatro}) and one extra condition which fixes
$\partial_r \alpha$ in eq. (\ref{ssiete}). It is thus natural to think that
the number of independent functions is six and, thus, in principle, one should
be able to express the metric and the BPS equations in terms of them. By
looking at the complicated form of the first-order equations and  constraints
one could be tempted to think that this is a hopeless task. However, we will
show that this is not the case and that there exists a set of variables,
which are precisely those introduced by Hitchin in ref. \cite{hit}, in which 
the BPS equations drastically
simplify. These equations involve the constants $p$ and $q$ just discussed,
together with the components of the calibrating three-form $\Phi$. Actually,
following refs. \cite{hit,bran,Chong}, we shall parametrize $\Phi$ as:
\beq
\Phi\,=\,e^{7}\,\wedge\,\omega\,+\,\rho\,\,,
\label{setenta}
\eeq
where the two-form $\omega$ is given in terms of three functions $y_i$ as:
\beq
\omega\,=\,\sqrt{{y_2y_3\over y_1}}\,w^1\wedge\tilde w^1\,+\,
\sqrt{{y_3y_1\over y_2}}\,w^2\wedge\tilde w^2\,+\,
\sqrt{{y_1y_2\over y_3}}\,w^3\wedge\tilde w^3\,,
\label{stuno}
\eeq
and $\rho$ is a three-form which depends on another set of three functions
$x_i$, namely:
\bear
\rho\,&=&\,p\,w^1\wedge w^2\wedge w^3\,+
\,q\,\tilde w^1\wedge \tilde w^2\wedge \tilde w^3\,+\rc\rc
&&+\,x_1\,\big(\, w^1\wedge \tilde w^2\wedge \tilde w^3\,-\,
w^2\wedge  w^3\wedge \tilde w^1\,\big)\,+\,{\rm cyclic}\,\,.
\label{stdos}
\eear
Notice that the terms appearing in $\omega$ are precisely those which follow
from our expression (\ref{cuno}) for $\Phi$. Moreover, by plugging on the
right-hand side of eq. (\ref{cuno}) the relation (\ref{socho}) between the
one-forms
$e^A$ and the SU(2) left invariant forms, one can find the explicit relation
between the new and old variables, namely:
\bear
y_1&=&e^{{2\phi\over 3}+h_2+h_3-\lambda_1}\,\,,\rc\rc
x_1&=&-2\,\big[\,e^{\phi+h_1-\lambda_1}\,\cos\alpha\,+\,
4\,e^{2\phi}\,G_1\sin\alpha\,\big]\,\,,
\label{sttres}
\eear
and cyclically in $(1,2,3)$. Notice that the coefficients of
$w^1\wedge \tilde w^2\wedge \tilde w^3$ and of
$-w^2\wedge  w^3\wedge \tilde w^1$ in the expression
(\ref{stdos}) of $\rho$ must be necessarily equal if $\Phi$ is closed.
Actually,
by computing the latter in our formalism, we get an alternative expression for
the $x_i$'s. This other expression  is:
\bear
x_1&=&\,2\,\big[\,e^{h_3-\lambda_3}\,G_2\,+\,e^{h_2-\lambda_2}\,G_3\,\big]\,
e^{\phi}\,\cos\alpha\,+\,\rc\rc
&&+\,\big[\,8\,e^{2\phi}\,G_2\,G_3\,-\,{1\over
2}\,e^{\lambda_1+h_2+h_3}\,\big]\,
\sin\alpha\,\,,
\label{stcuatro}
\eear
and cyclically in $(1,2,3)$. As a matter of fact, these 
two alternative expressions for the $x_i$'s are equal as a consequence of
the constraints  (\ref{suno}). In fact, we can regard eqs. (\ref{suno})
and (\ref{ssiete}) as conditions needed to ensure the
closure of $\Phi$. On the other hand, by using, at our convenience, eqs.
(\ref{sttres}) and (\ref{stcuatro}), one can prove the  following useful
relations:
\bear
&&{x_2x_3-px_1\over y_1}\,=\,{1\over 4}\,e^{2h_1-{2\phi\over 3}}\,+\,
4\,e^{{4\phi\over 3}+2\lambda_1}\,\,G_1^2\,\,,\rc\rc
&&{x_1^2-x_2^2-x_3^2-pq\over y_1}\,=\,
8\,e^{{4\phi\over 3}+2\lambda_1}\,\,G_1\,\,,\rc\rc
&&{x_2x_3+qx_1\over y_1}\,=\,4\,e^{{4\phi\over 3}+2\lambda_1}\,\,,
\label{stcinco}
\eear
and cyclically in $(1,2,3)$. As a first application of eq. (\ref{stcinco}),
let us point out that, making use of this equation, one can easily invert the
relation (\ref{sttres}). The result is:
\bear
e^{2\phi}\,&=&\,{1\over 8}\,\,
{(qx_1+x_2x_3)^{1/2}(qx_2+x_1x_3)^{1/2}(qx_3+x_1x_2)^{1/2}
\over \sqrt{y_1y_2y_3}}\,\,,\rc\rc\rc
e^{2\lambda_1}\,&=&\,{(y_2y_3)^{1/3}\over (y_1)^{2/3}}\,\,
{(qx_1+x_2x_3)^{2/3}
\over (qx_2+x_1x_3)^{1/3}(qx_3+x_1x_2)^{1/3}}\,\,,\rc\rc\rc
e^{2h_1}\,&=&\,2\,\,{(y_2y_3)^{5/6}\over (y_1)^{1/6}}\,\,
{(qx_2+x_1x_3)^{1/6}(qx_3+x_1x_2)^{1/6}
\over (qx_1+x_2x_3)^{5/6}}\,\,,\rc\rc\rc
G_1&=&{1\over 2}\,\,{x_1^2-x_2^2-x_3^2-pq\over
qx_1+x_2x_3}\,\,,
\label{stseis}
\eear
and cyclically in $(1,2,3)$.
Moreover, in order to make  contact with the formalism of refs. \cite{hit,
Chong},  let us define now the following ``potential":
\bear
U&\equiv&p^2q^2\,+\,2pq\,(\,x_1^2\,+\,x_2^2\,+\,x_3^2\,)\,+\,
4(p-q)\,x_1x_2x_3\,+\,\rc\rc
&&+\,x_1^4\,+\,x_2^4\,+\,x_3^4\,-\,2x_1^2x_2^2\,
-\,2x_2^2x_3^2\,-\,2x_3^2x_1^2\,\,.
\label{stsiete}
\eear
A straightforward calculation shows that $U$ can be rewritten as:
\bear
U&=&{1\over 3}\,(\,x_1^2\,-\,x_2^2\,-\,x_3^2\,-\,pq\,)^2\,-\,
{4\over 3}\,(\,x_2x_3\,+\,qx_1\,)\,(\,x_2x_3\,-\,px_1\,)\,+\,\rc\rc
&&+\,{\rm cyclic \,\,\,permutations}\,\,.
\label{stocho}
\eear
By using (\ref{stcinco}) to evaluate the right-hand side of eq.
(\ref{stocho}),
together with the definition of the $y_i$'s written in eq. (\ref{sttres}), one
easily verifies that $U$ is given by:
\beq
U\,=\,-4y_1y_2y_3\,\,.
\label{stnueve}
\eeq
It is important to stress the fact that in the general Hitchin formalism the
relation (\ref{stnueve}) is a constraint, whereas here this equation is just
an identity which follows from the definitions of $p$, $q$, $x_i$ and $y_i$.
Another
important consequence of the identities (\ref{stcinco}) is the form of the
metric in the new variables. Indeed, it is immediate from eqs.
(\ref{socho}) and (\ref{stcinco}) to see that the seven dimensional metric
$ds^2_{7}$ takes the form:
\bear
&&ds^2_{7}=dt^{2}\,+\,\rc\rc
&&+\,{1\over y_1}\,\Big[\,\big(\,x_2x_3-px_1\,\big)\,(w^1)^2\,+\,
\big(\,x_1^2-x_2^2-x_3^2-pq\,\big)\,w^1\tilde w^1\,+\,
\big(\,x_2x_3+qx_1\,\big)\,(\tilde w^1)^2\,\Big]\,+\rc\rc
&&+\,{1\over y_2}\,\Big[\,\big(\,x_3x_1-px_2\,\big)\,(w^2)^2\,+\,
\big(\,x_2^2-x_3^2-x_1^2-pq\,\big)\,w^2\tilde w^2\,+\,
\big(\,x_3x_1+qx_2\,\big)\,(\tilde w^2)^2\,\Big]\,+\rc\rc
&&+\,{1\over y_3}\,\Big[\,\big(\,x_1x_2-px_3\,\big)\,(w^3)^2\,+\,
\big(\,x_3^2-x_1^2-x_2^2-pq\,\big)\,w^3\tilde w^3\,+\,
\big(\,x_1x_2+qx_3\,\big)\,(\tilde w^3)^2\,\Big]\,\,,\rc
\label{ochenta}
\eear
where $dt^2\,=\,e^{-2\phi/3}\,dr^2$.

It remains to determine the first-order system of differential equations
satisfied by the new variables. First of all, recall that, in the old
variables,  the BPS equations depend on the phase $\alpha$. Actually, from the
expression of $q$ (eq. (\ref{snueve})), and the first equation in
(\ref{stseis}), one can easily determine $\sin\alpha$, whereas $\cos\alpha$
can be obtained from the second equation in (\ref{sttres}). The result is:
\bear
\sin\alpha&=&-q\,\,
{\sqrt{y_1y_2y_2}\over
(qx_1+x_2x_3)^{1/2}(qx_2+x_1x_3)^{1/2}(qx_3+x_1x_2)^{1/2}}\,\,\,,\rc\rc\rc
\cos\alpha&=&-\,\,
{2x_1x_2x_3\,+\,q\,(x_1^2+x_2^2+x_3^2\,)\,+\,pq^2\over
2(qx_1+x_2x_3)^{1/2}(qx_2+x_1x_3)^{1/2}(qx_3+x_1x_2)^{1/2}}\,\,.
\label{ouno}
\eear
As a check of eq. (\ref{ouno}) one can easily verify that
$\sin^2\alpha+\cos^2\alpha=1$ as a consequence of the relation
(\ref{stnueve}).
It is now straightforward to compute the derivatives of $x_i$ and $y_i$.
Indeed, one can differentiate eq. (\ref{sttres}) and use eqs.
(\ref{cicinco}), (\ref{ciseis}), (\ref{cinueve}), (\ref{scinco}) and
(\ref{ssiete}) to evaluate the result in the old variables. This result can be
converted back to the new variables by means of eqs. (\ref{stseis}) and
(\ref{ouno}). The final result of these calculations is remarkably simple,
namely:
\bear
\dot x_1&=&-\sqrt{{y_2y_3\over y_1}}\,\,,\rc\rc
\dot y_1&=&{pqx_1\,+\,(p-q)x_2x_3\,+\,x_1(x_1^2-x_2^2-x_3^2)
\over \sqrt{y_1y_2y_3}}\,\,,
\label{odos}
\eear
and cyclically in $(1,2,3)$. In eq. (\ref{odos}) the dot denotes derivative
with respect to the variable $t$ defined after eq. (\ref{ochenta}). The
first-order system (\ref{odos}) is, with our notations, the one derived in
refs. \cite{hit, Chong}. Indeed, one can show that the equations satisfied
by the $x_i$'s are a consequence of the condition $d\Phi=0$, whereas, if
the seven dimensional Hodge
dual is computed with the metric (\ref{ochenta}), then $d*_7\Phi=0$ implies
the
first-order equations for the $y_i$'s. Therefore, we have shown that eight
dimensional gauged supergravity provides an explicit realization of the
Hitchin
formalism for general values of the constants $p$ and $q$. Notice that a
non-zero phase $\alpha$ is needed in order to get a system with $q\not=0$.
Recall (see eq. (\ref{ttres})) that the phase $\alpha$ parametrizes the
tilting of the three cycle on which the D6-brane is wrapped with respect
to the three sphere of the eight dimensional metric. Notice that the
analysis of \cite{hs1} corresponds to the case $q = \alpha = 0$.

Let us finally point out that the first-order equations  (\ref{odos})  are
invariant if we change the constants $(p,q)$ by $(-q,-p)$. In the metric
(\ref{ochenta}) this change is equivalent to the exchange of $w^i$ and
$\tilde w^i$, \ie\ of the two $S^3$ of the principal orbits of the
cohomogeneity
one metric (\ref{ochenta}). As mentioned above, this is the so-called flop
transformation. Thus, we have proved that:
\beq
w^i\leftrightarrow\tilde w^i\,\,\,\,
\Longleftrightarrow\,\,\,\,(p,q)\,\rightarrow (-q,-p)\,\,.
\label{otres}
\eeq
Notice that the three-form $\Phi$ given in eqs. (\ref{setenta})-(\ref{stdos})
changes its sign when both $(w^i,\tilde w^i)$ and $(p,q)$ are transformed
as in eq. (\ref{otres}).

\setcounter{equation}{0}
\section{Some particular cases}
\medskip

With the kind of ansatz we are adopting for the eight-dimensional
solutions, the
corresponding  eleven dimensional metrics are of  the type:
\beq
ds^2_{11}\,=\,dx^2_{1,3}\,+\,B_i^2\,\,
(\,w^i\,)^2\,+\,D_i^2\,
(\,\tilde w^i\,+\,G_i\,w^i\,)^2\,+\,dt^2\,\,,
\label{ocuatro}
\eeq
where the coefficients $B_i$, $D_i$ and the variable $t$ are related to eight
dimensional quantities as follows:
\beq
B_i^2\,=\,{1\over 4}\,e^{2h_i\,-\,{2\phi\over 3}}\,\,,
\,\,\,\,\,\,\,\,\,\,\,\,\,\,\,\,
D_i^2\,=\,4\,e^{{4\phi\over 3}+2\lambda_i}\,\,,
\,\,\,\,\,\,\,\,\,\,\,\,\,\,\,\,
dt^2\,=\,e^{-{2\phi\over 3}}\,dr^2\,\,.
\label{ocinco}
\eeq
Moreover, we have found that, for a supersymmetric solution, the nine
functions appearing in the metric are not independent but rather they are
related by some algebraic constraints which are, in general, quite
complicated. Notice that, in this case, the gauged supergravity approach
forces the six function ansatz, this possibly clarifying the reasons
behind this {\em a priori} requirement in previous cases in the literature.
To illustrate
this point, let us write eq. (\ref{scuatro}) in terms of $B_i$, $D_i$ and
$G_i$. One gets:
\bear
&&\big[\,B_1\,D_2^2\,D_1\,G_2\,-\,(\,1\leftrightarrow 2\,)\,\big]\,D_3^2\,
(\,1\,-\,G_3^2\,)\,=\,\rc\rc
&&=\,B_3\,D_2\,\big[\,B_1\,B_3\,D_1\,D_2\,G_2\,+\,
D_1^2\,D_2\,D_3\,G_1^2\,+\,B_1^2\,D_2\,D_3\,\big]\,-\,
(\,1\leftrightarrow 2\,)\,\,,
\label{oseis}
\eear
and cyclically in $(1,2,3)$. In addition, we must ensure that eq.
(\ref{ssiete}) is also satisfied. Despite the terrifying aspect of eq.
(\ref{oseis}), it is not hard to find expression for, say, $G_2$ and $G_3$
in terms of the remaining functions. Moreover, we will be able to find
some particular solutions, which
correspond to the different cohomogeneity one metrics with $S^3\times S^3$
principal orbits and SU(2)$\times$ SU(2) isometry which have been studied in
the literature.

\subsection{The q=0 solution}
\medskip
The simplest way of solving the constraints imposed by supersymmetry is by
taking $q=0$. A glance at the second equation in (\ref{snueve}) reveals that
in this case $\sin\alpha=0$ and, thus, $\beta=1$, $\tilde\beta=0$. Notice,
first of all, that this is a consistent solution of eq. (\ref{ssiete}).
Moreover, it follows from eq. (\ref{suno}) that one must have:
\beq
{\cal A}_i\,=\,0\,\,.
\label{osiete}
\eeq
By combining the three conditions (\ref{osiete}) it is easy to find the values
of the gauge field components $G_i$ in terms of the other functions $B_i$ and
$D_i$ \cite{hs1}. One gets:
\beq
G_1\,=\,{1\over 2}\,\,{D_2D_3\over B_2B_3}\,\,
\Big[\,\,\Big({B_1\over D_1}\Big)^2\,-\,\Big({B_2\over D_2}\Big)^2\,-\,
\Big({B_3\over D_3}\Big)^2\,\,\Big]\,\,,
\label{oocho}
\eeq
and cyclically in $(1,2,3)$, which is precisely the result of \cite{hs1}.
This is the solution of the constraints we were
looking for. One can check that, assuming that the $G_i$'s are given by  eq.
(\ref{oocho}), then eq.  (\ref{scinco}) for $G_i'$  is satisfied if eqs.
(\ref{cicinco}), (\ref{ciseis}) and (\ref{cinueve})  hold. Thus, eq.
(\ref{oocho}) certainly gives a consistent truncation of the first-order
differential equations. On the other hand, by using the value of the $G_i$'s
given in eq.  (\ref{oocho}), one can eliminate them and obtain a system of
first-order equations for the remaining functions $B_i$ and $D_i$. These
equations are:
\bear
\dot B_1&=&-{D_2\over 2B_3}\,(\,G_2\,+\,G_1G_3\,)\,-\,
{D_3\over 2B_2}\,(\,G_3\,+\,G_1G_2)\,\,,\rc\rc
\dot D_1&=&{D_1^2\over 2B_2B_3}\,(\,G_1\,+\,G_2G_3\,)\,+\,
{1\over 2D_2D_3}\,(\,D_2^2\,+\,D_3^2\,-\,D_1^2\,)\,\,,
\label{onueve}
\eear
together with the other permutations of the indices $(1,2,3)$. In
(\ref{onueve}) the $G_i$'s are are the functions of $B_i$ and $D_i$
displayed in
eq. (\ref{oocho}). The constant $p$ can be immediately obtained from
(\ref{snueve}), namely:
\beq
p=B_1B_2B_3\,-\,B_1D_2D_3G_2G_3\,-\,B_2D_1D_3G_1G_3\,-\,B_3D_1D_2G_1G_2\,\,.
\label{noventa}
\eeq

Let us now give the Hitchin variables in this case. By taking $\alpha=0$ on
the right-hand side of (\ref{sttres}) and using the relation (\ref{ocinco}),
one readily arrives at:
\beq
x_1\,=\,-B_1D_2D_3\,\,,
\,\,\,\,\,\,\,\,\,\,\,\,\,\,\,\,
y_1\,=\,B_2B_3D_2D_3\,\,.
\label{nuno}
\eeq
The values of the other $x_i$ and $y_i$ are obtained by cyclic permutation. As
a verification of these expressions, it is not difficult to demonstrate, by
using eq. (\ref{onueve}), that the functions $x_i$ and $y_i$ of
eq. (\ref{nuno}) satisfy the first-order equations (\ref{odos}) for $q=0$.
Finally, let us point out that, by means of a flop transformation, one
can pass from the $q=0$ metric
described above to a metric with $p=0$.

\subsection{The flop invariant solution}
\medskip

It is also possible  to solve our constraints by requiring that the metric be
invariant under the $\ZZ_2$ flop transformation $w^i\leftrightarrow \tilde
w^i$. It follows from eq. (\ref{otres}) that, in this case, we must
necessarily have $p=-q$. Moreover, it is also clear that the forms $w^i$
and $\tilde w^i$ must enter the metric in the combinations
$(\,w^i+\tilde w^i\,)^2$ and $(\,w^i-\tilde w^i\,)^2$, which are the
only quadratic combinations which
are invariant under the flop transformation. Thus the metric we are seeking
must be of the type:
\beq
ds^2_{11}\,=\,dx^2_{1,3}\,+\,a_i^2\,(\,w^i\,-\,\tilde w^i\,)^2\,+\,
b_i^2\,(\,w^i\,+\,\tilde w^i\,)^2\,+\,dt^2\,\,,
\label{ndos}
\eeq
where $a_i$ and $b_i$ are functions which obey some system of first-order
differential equations to be determined. In general \cite{CGLP1}, a metric
of the type written in eq. (\ref{ocuatro}) can be put in the form
(\ref{ndos}) only if
$G_i$, $B_i$ and $D_i$ satisfy the following relation:
\beq
G_i^2\,=\,1\,-\,{B_i^2\over D_i^2}\,\,.
\label{ntres}
\eeq
It is easy to show that our constraints are solved for $G_i$ given as in eq.
(\ref{ntres}). Indeed, after some calculations, one can rewrite the constraint
(\ref{scuatro}) for $i=1$ and $j=2$  as:
\bear
&&\Big(\,1\,-\,{1\over 16}\,e^{-2\phi+2h_1-2\lambda_1}\,-\,G_1^2\,\Big)\,
\,e^{-2\lambda_3}\,-\,
\Big(\,1\,-\,{1\over 16}\,e^{-2\phi+2h_2-2\lambda_2}\,-\,G_2^2\,\Big)\,
\,e^{-2\lambda_3}\,+\,\rc\rc
&&+\,\Big(\,1\,-\,{1\over 16}\,e^{-2\phi+2h_3-2\lambda_3}\,-\,G_3^2\,\Big)\,
\Big[\,G_2\,e^{h_1-h_3+\lambda_2}\,-\,
G_1\,e^{h_2-h_3+\lambda_1}\,\Big]\,\,=0\,\,,
\label{ncuatro}
\eear
which is clearly solved for:
\beq
G_i^2\,=\,1\,-\,{1\over 16}\,e^{-2\phi+2h_i-2\lambda_i}\,\,.
\label{ncinco}
\eeq
Similarly, one can verify that eq. (\ref{ncinco}) also solves eq.
(\ref{scuatro}) for the remaining values of $i$ and $j$. After taking into
account the identifications (\ref{ocinco}), we easily conclude that the
solution (\ref{ncinco}) coincides with the condition  (\ref{ntres}) and,
thus, it corresponds to $\ZZ_2$ invariant metric of the type (\ref{ndos}).
Moreover, it
can be checked that the relation (\ref{ncinco}) gives a consistent
truncation of
the first-order differential equations found in section 3 and that eq.
(\ref{ssiete}) is also satisfied. On the other hand, the identification 
of the $a_i$ and $b_i$ functions with the ones corresponding to 8d gauged
supergravity is easily established by comparing the uplifted metric with
(\ref{ndos}), namely:
\bear
dr&=&e^{{\phi\over 3}}\,dt\,\,,\rc\rc
e^{2h_i-{2\phi\over 3}}&=&16\,{a_i^2\,b_i^2\over
a_i^2\,+\,b_i^2}\,\,,\rc\rc
e^{{4\phi\over 3}+2\lambda_i}&=&{1\over 4}\,(\,a_i^2\,+\,b_i^2\,)\,\,.
\label{nseis}
\eear
This relation allows to obtain $\phi$, $\lambda_i$ and $h_i$ in terms of
$a_i$ and $b_i$:
\bear
e^{2\phi}\,&=&\,{1\over 8}\,\,\prod_i\,
(\,a_i^2\,+\,b_i^2)^{1\over 2}\,\,,\rc\rc
e^{2\lambda_i}\,&=&\,{a_i^2\,+\,b_i^2\over
\prod_j\,(\,a_j^2\,+\,b_j^2)^{1\over 3}}\,\,,\rc\rc
e^{2h_i}\,&=&\,8\,{a_i^2\,b_i^2\over a_i^2\,+\,b_i^2}\,\,
\prod_j\,(\,a_j^2\,+\,b_j^2)^{1\over 6}\,\,,
\label{nsiete}
\eear
while $G_i$ in terms of the $a_i$ and $b_i$ is given by:
\beq
G_i\,=\,{b_i^2-a_i^2\over  b_i^2+a_i^2}\,\,.
\label{nocho}
\eeq
The inverse relation is also useful:
\beq
a_i^2\,=\,2\,e^{{4\phi\over 3}\,+\,2\lambda_i}\,\,
(\,1\,-\,G_i\,)\,\,,
\,\,\,\,\,\,\,\,\,\,\,\,\,
b_i^2\,=\,2\,e^{{4\phi\over 3}\,+\,2\lambda_i}\,\,
(\,1\,+\,G_i\,)\,\,,
\label{nnueve}
\eeq
where $G_i$ is the function of $\phi$, $h_i$ and $\lambda_i$ written in eq.
(\ref{ncinco}). By using eqs. (\ref{nsiete}) and (\ref{nocho}) one can obtain
the values of $\cos\alpha$ and $\sin\alpha$ for this case. One gets:
\bear
\cos\alpha&=&
{\,b_1a_2a_3\,+\,a_1b_2a_3\,+\,a_1a_2b_3\,-\,b_1b_2b_3\,\over
\sqrt{(\,a_1^2\,+\,b_1^2\,)\,(\,a_2^2\,+\,b_2^2)\,(\,a_3^2\,+\,b_3^2\,)}}
\,\,,\rc\rc
\sin\alpha&=&
{\,a_1b_2b_3\,+\,b_1a_2b_3\,+\,b_1b_2a_3\,-\,a_1a_2a_3\over
\sqrt{(\,a_1^2\,+\,b_1^2\,)\,(\,a_2^2\,+\,b_2^2)\,(\,a_3^2\,+\,b_3^2\,)}}
\,\,.
\label{cien}
\eear
Moreover, by differentiating eq. (\ref{nnueve}) and using the first-order
equations of section 3, together with eqs. (\ref{nsiete}) and (\ref{cien}),
one can find the BPS equations in the $a_i$ and $b_i$ variables. They are:
\bear
\dot a_1&=&
-\,{a_1^2\over 4a_2b_3}\,-\,{a_1^2\over 4a_3b_2}\,+\,
{a_2\over 4b_3}\,+\,{b_2\over 4a_3}\,+\,
{a_3\over 4b_2}\,+\,{b_3\over 4a_2}\,\,,\rc\rc
\dot b_1&=&
-\,{b_1^2\over 4a_2a_3}\,+\,{b_1^2\over 4b_2b_3}\,-\,
{b_2\over 4b_3}\,+\,{a_2\over 4a_3}\,-\,
{b_3\over 4b_2}\,+\,{a_3\over 4a_2}\,\,,
\label{ctuno}
\eear
and cyclically for the other $a_i$'s and $b_i$'s. These are precisely the
equations found in ref. \cite{bggg}  for this type of metrics. Moreover, it
is now
straightforward to compute the constants $p$ and $q$ in this case. Indeed, by
substituting eqs. (\ref{nsiete}), (\ref{nocho}) and (\ref{cien}) on the
right-hand side of eq. (\ref{snueve}), one easily proves that:
\beq
p\,=\,-q\,=\,\,a_1b_2b_3\,+\,b_1a_2b_3\,+\,b_1b_2a_3\,-\,a_1a_2a_3\,\,.
\label{ctdos}
\eeq
Similarly, from eq. (\ref{sttres}) one can find
the Hitchin variables in terms of the $a_i$'s and $b_i$'s. The result for
$x_1$ and $y_1$ is:
\bear
x_1&=&a_1b_2b_3\,-\,b_1a_2b_3\,-\,b_1b_2a_3\,-\,a_1a_2a_3\,\,,\rc\rc
y_1&=&4a_2a_3b_2b_3\,\,,
\label{cttres}
\eear
while the expressions of $x_2$, $x_3$, $y_2$ and $y_3$ are obtained from
(\ref{cttres}) by cyclic permutations.

\subsection{The conifold-unification metrics}
\medskip

There exist a class of $G_2$ metrics with $S^3\times S^3$ principal orbits
which have an extra $U(1)$ isometry and generic values of $p$ and $q$.
They are the
so-called conifold--unification metrics and they were introduced in
ref. \cite{CGLP2} as a
unification, via M--theory, of the deformed and resolved conifolds. Following
ref. \cite{CGLP2}, let us parametrize them as:
\bear
ds^2_7&=&a^2\,\big[\,\big(\,\tilde w^1\,+\,
{\cal G}\,w^1\,\big)^2\,+\,
\big(\,\tilde w^2\,+\,
{\cal G}\,w^2\,\big)^2\,\big]\,+\,
b^2\,\big[\,\big(\,\tilde w^1\,-\,
{\cal G}\,w^1\,\big)^2\,+\,
\big(\,\tilde w^2\,-\,
{\cal G}\,w^2\,\big)^2\,\big]\,+\rc\rc
&&+\,c^2\,\big(\,\tilde w^3\,-\,w^3\,\big)^2\,+\,
f^2\,\big(\,\tilde w^3\,+\,{\cal G}_3\,w^3\,\big)^2\,+\,dt^2\,\,.
\label{ctcuatro}
\eear
It is clear that, in order to obtain in our eight-dimensional supergravity
approach a metric such as the one written in eq. (\ref{ctcuatro}), one must
take  $h_1=h_2$, $\lambda_1=\lambda_2\,=\,-\lambda_3/2\,=\,\lambda$ and
$G_1=G_2$ in our general formalism. Then, it is an easy exercise to find the
gauged supergravity variables in terms of the functions appearing in the
ansatz (\ref{ctcuatro}). One has:
\bear
&&e^{\phi}\,=\,{1\over 2\sqrt{2}}\,\big(\,a^2\,+\,b^2\,\big)^{{1\over 2}}\,\,
\big(\,f^2\,+\,c^2\,\big)^{{1\over 4}}\,\,,\rc\rc
&&e^{\lambda}\,=\,\big(\,a^2\,+\,b^2\,\big)^{{1\over 6}}\,\,
\big(\,f^2\,+\,c^2\,\big)^{-{1\over 6}}\,\,,\rc\rc
&&e^{h_1}\,=\,2\sqrt{2}\,a\,b\,{\cal G}\,
\big(\,a^2\,+\,b^2\,\big)^{-{1\over 3}}\,\,
\big(\,f^2\,+\,c^2\,\big)^{{1\over 12}}\,\,,\rc\rc
&&e^{h_3}\,=\,\sqrt{2}\,f\,c\,(1+{\cal G}_3)\,
\big(\,a^2\,+\,b^2\,\big)^{{1\over 6}}\,\,
\big(\,f^2\,+\,c^2\,\big)^{-{5\over 12}}\,\,,\rc\rc
&&G_1\,=\,{\cal G}\,\,{a^2-b^2\over a^2+b^2}\,\,,\rc\rc
&&G_3\,=\,\,{{\cal G}_3\,\,f^2-c^2\over f^2+c^2}\,\,.
\label{ctcinco}
\eear
With the parametrization given above, it is not difficult to solve the
constraints (\ref{scuatro}). Actually, one of these constraints is trivial,
while the other allows to obtain ${\cal G}_3$ in terms of the other variables,
namely:
\beq
{\cal G}_3\,=\,{\cal G}^2\,+\,{c\,(\,a^2-b^2\,)(\,1-{\cal G}^2\,)
\over 2abf}\,\,.
\label{ctseis}
\eeq
The relation (\ref{ctseis}), with $a\rightarrow -a$, is precisely the one
obtained in ref. \cite{CGLP2}. One can also prove that eq. (\ref{ctseis})
solves
eq.  (\ref{ssiete}). Actually, the phase $\alpha$ in this case is:
\beq
\cos\alpha\,=\,{2abc\,+\,(b^2-a^2)\,f\over
(\,a^2+b^2\,)\sqrt{c^2+f^2}}\,\,,
\,\,\,\,\,\,\,\,\,\,\,\,\,\,\,\,\,\,
\sin\alpha\,=\,{2abf\,+\,(a^2-b^2)\,c\over
(\,a^2+b^2\,)\sqrt{c^2+f^2}}\,\,.
\label{ctsiete}
\eeq

With all these ingredients  it is now straightforward, although tedious, to
find the first-order equations for the five independent functions of the
ansatz
(\ref{ctcuatro}). The result coincides again with the one written in
ref.\cite{CGLP2}, after changing $a\rightarrow -a$, and is given by:
\bear
\dot a&=&{c^2\,(\,b^2-a^2\,)\,+\,[\,4a^2\,(\,b^2-a^2\,)\,+\,
c^2\,(\,5a^2-b^2\,)\,-\,4abcf\,]\,{\cal G}^2\over
16a^2\,bc\,{\cal G}^2}\,\,,\rc\rc
\dot b&=&{c^2\,(\,a^2-b^2\,)\,+\,[\,4b^2\,(\,a^2-b^2\,)\,+\,
c^2\,(\,5b^2-a^2\,)\,+\,4abcf\,]\,{\cal G}^2\over
16ab^2\,c\,{\cal G}^2}\,\,,\rc\rc
\dot c&=&-{c^2\,+\,(\,c^2\,-\,2a^2\,-\,2b^2\,)\,{\cal G}^2\over
4ab{\cal G}^2}\,\,,\rc\rc
\dot f&=&-{(\,a^2-b^2\,)\,\big[\,4abf^2\,{\cal G}^2\,+\,
c\,(\,a^2\,f\,-\,b^2\,f\,-\,4abc\,)\,(\,1-{\cal G}^2\,)\,\big]\over
16a^3\,b^3\,{\cal G}^2}\,\,,\rc\rc
\dot {\cal G}&=&{c\,(\,1-{\cal G}^2\,)\over 4ab{\cal G}}\,\,.
\label{ctocho}
\eear
Furthermore, the constants $p$ and $q$ are also easily obtained, with the
result:
\bear
p&=&(\,a^2\,-\,b^2\,)\,c\,{\cal G}^2\,+\,
2abf{\cal G}_3\,{\cal G}^2\,\,,\rc\rc
q&=&(\,b^2\,-\,a^2\,)\,c\,-\,2abf\,\,,
\label{ctnueve}
\eear
while the Hitchin variables are:
\bear
&&x_1\,=\,x_2\,=\,-(\,a^2+b^2\,)\,c\,{\cal G}\,\,,
\,\,\,\,\,\,\,\,\,\,\,\,\,\,\,\,\,\,
x_3\,=\,(\,a^2-b^2\,)\,c\,-2abf{\cal G}_3\,\,,\rc\rc
&&y_1\,=\,y_2\,=\,2abcf{\cal G}(\,1\,+\,{\cal G}_3\,)\,\,,
\,\,\,\,\,\,\,\,\,\,\,\,\,\,\,\,\,\,
y_3\,=\,4a^2b^2{\cal G}^2\,\,.
\label{ctdiez}
\eear
Eqs. (\ref{ctnueve}) and (\ref{ctdiez}) are again in agreement with
those given in ref. \cite{CGLP2}, after changing $a$ by $-a$ as before.

\setcounter{equation}{0}
\section{Summary and Conclusions}
\medskip

In this paper we have studied the supersymmetric configurations of eleven
dimensional supergravity which are the direct product of Minkowski four
dimensional spacetime and a cohomogeneity one seven dimensional manifold of
$G_2$ holonomy with $S^3\times S^3$ principal orbits. These configurations
are obtained by uplifting to eleven dimensions
some solutions of eight dimensional gauged supergravity which preserve four
supersymmetries and satisfy a system of first-order BPS equations. They can be
interpreted as being originated by D6-branes wrapping a supersymmetric three
cycle which corresponds to a domain wall in eight dimensional gauged
supergravity.

The supersymmetry of the solutions is guaranteed by the BPS equations which,
once a careful adjustment of the spin connection and the $SU(2)$ gauge field
of the eight--dimensional
theory has been made,  are the conditions required to have Killing spinors.
This adjustment is what is known as the topological twist and is directly
related to the projection conditions imposed to the Killing spinors. In this
paper we have shown how to generalize this projection with respect to the one
used up to now. This generalization amounts to the introduction of a phase
$\alpha$ in the radial projection of the Killing spinor and,
correspondingly, the
twist is implemented by a non-abelian gauge field which is not fixed a priori
(as in the  previous approaches in the literature) but determined by a
first-order differential equation. This gauge field encodes the non trivial
fibering of the two three spheres in the special holonomy manifold, while the
corresponding radial projection determines the wrapping of the D6-brane in the
supersymmetric three cycle. Actually we have seen that, for non-zero $\alpha$,
the three cycle on which the D6-brane is wrapped has components along the two
$S^3$'s (see eq. (\ref{ttres})).

A careful analysis of the conditions imposed by supersymmetry has revealed us
that, for a general ansatz as in eqs. (\ref{csiete})--(\ref{cnueve}), some
algebraic constraints have to be imposed to the functions of the ansatz.  We
have verified that all metrics studied in the literature are particular
solutions of our constraints and, in fact, we have found a map between our
system and the one introduced by Hitchin. In particular we have demonstrated
that, contrary to the generalized believe,  the metrics with $q\not= 0$ can be
obtained within the 8d gauged supergravity approach. Our formalism is general
and systematic and does not assume any particular form of the 
seven--dimensional metric.

There are other instances on which the kind of generalized twist introduced
here can also be studied, the most obvious of them being the cases of
D6--branes wrapping two and four cycles. In the former situation we would have
to deal with Calabi-Yau manifolds, whereas when the D6--branes wrap a four
cycle the special holonomy manifold would be eight dimensional and would have
Spin(7) or $SU(4)$ holonomy depending on whether the cycle is coassociative
or K\"ahler. This would be a powerful technique to seek for complete metrics
for these special holonomy manifolds. It would be also interesting to analyze
the ten dimensional supergravity solutions which correspond to fivebranes
wrapping two and three cycles. The relevant gauged supergravity for these
cases lives in seven dimensions. Actually, an implementation of the twisting
similar to the one introduced here was used in \cite{abcpz} to obtain the
Maldacena--N\'u\~nez solution \cite{mn2} for the supergravity dual of
${\cal N} = 1$ super Yang--Mills theory.

It would be interesting to study the effect of turning on fluxes in this
framework, extending previous results in refs. \cite{gns,epr1}. The
generalization of the twisting seems general enough so as to deserve a
more careful study in many lower dimensional gauged supergravities. In
particular, it would be interesting to seek for more solutions that do
not correspond to the near horizon limit of wrapped D--branes.
It is intriguing, for example, to see whether or not the full
flat D--brane solution ({\rm i.e.} without the near horizon limit being
taken on it, such as, for example, the Taub--NUT metric for the
D6--brane) can be obtained within lower dimensional gauged supergravity.

We are currently working on these issues and we hope to report on
them in a near future.

\medskip
\section*{Acknowledgments}
\medskip
We are grateful to Jaume Gomis, Joaquim Gomis, Rafael Hernandez,
Carlos N\'u\~nez and Javier Mas for highly useful discussions and
comments that enriched this article. AVR would like to thank the
IST for kind hospitality during the final stages of this paper.
This work has been supported
in part by MCyT and FEDER under grant  BFM2002-03881, by Xunta de
Galicia, by Fundaci\'on Antorchas and by Funda\c c\~ao para a
Ci\^encia e a Tecnologia under grants POCTI/1999/MAT/33943 and
SFRH/BPD/7185/2001.

\vskip 1cm
\renewcommand{\theequation}{\rm{A}.\arabic{equation}}
\setcounter{equation}{0}
\medskip
\appendix
\section{D=8 gauged supergravity}
\medskip

The maximal gauged supergravity in eight dimensions was obtained in ref.
\cite{ss} by means of a Scherk-Schwarz \cite{ssch} reduction of eleven
dimensional supergravity on a SU(2) group manifold. In the bosonic sector
the field content of this theory includes the metric $g_{\mu\nu}$, a
dilatonic scalar $\phi$, five scalars parametrized by a $3\times 3$
unimodular matrix $L_i^{\alpha}$ which takes values in the coset
${\rm SL}(3,\RR)/{\rm SO}(3)$ and a SU(2) gauge potential  $A_{\mu}^i$.
In the fermionic sector there are two
pseudo Majorana spinors $\psi_{\mu}$ (the gravitino) and $\chi_i$ (the
dilatino). The kinetic energy of the coset scalars $L_i^{\alpha}$ is given in
terms of the symmetric traceless matrix $P_{\mu\,ij}$, defined through the
expression:
\beq
P_{\mu\,ij}+Q_{\mu\,ij}\,\,=\,L_i^{\alpha}\,
(\,\partial_{\mu}\,\delta_{\alpha}^{\beta}\,-\,
\epsilon_{\alpha\beta\gamma}\,A_{\mu}^{\gamma}\,)\,L_{\beta j}\,\,,
\label{apauno}
\eeq
where $Q_{\mu\,ij}$ is defined as the antisymmetric part of the right-hand
side of eq. (\ref{apauno}). For convenience we are setting in (\ref{apauno}),
and in what follows, the  SU(2)  coupling constant to one. Moreover, the
potential energy of the coset scalars is governed by the so-called
$T$-tensor, $T^{ij}$, and by its trace $T$, which are defined as:
\beq
T^{ij}\,=\,L^i_{\alpha}\,L^j_{\beta}\,\delta^{\alpha\beta}\,\,,
\,\,\,\,\,\,\,\,\,\,\,\,\,\,\,\,\,\,\,\,\,\,\,\,
T\,=\,\delta_{ij}\,T^{ij}\,\,.
\label{apados}
\eeq
Let $F_{\mu\nu}^{i}$ denote the field strength of the SU(2) gauge potential
$A_{\mu}^i$. Then, the lagrangian for the bosonic fields listed above is:
\bear
{\cal L}&=&\sqrt{-g_{(8)}}\,\,\Big[\,
{1\over 4}\,R\,-\,{1\over 4}\,e^{2\phi}\,
F_{\mu\nu}^{i}\,F^{\mu\nu\,\,i}\,-\,{1\over 4}\,
P_{\mu\,ij}\,P^{\mu\,ij}\,-\,{1\over 2}\,
\partial_{\mu}\,\phi\partial^{\mu}\,\phi\,-\cr\cr
&&-{1\over 16}\,e^{-2\phi}\,(\,T_{ij}\,T^{ij}\,-\,{1\over 2}T^2\,)
\,\,\Big]\,\,.
\label{apatres}
\eear
For any solution of the equations of motion derived from (\ref{apatres}), one
can write an eleven dimensional metric which solves the equations of $D=11$
supergravity. The corresponding uplifting formula is:
\beq
ds^2_{11}\,=\,e^{-{2\over 3}\,\phi}\,ds^2_8\,+\,4\,e^{{4\over 3}\phi}\,
(\,A^i\,+\,{1\over 2}\,L^i\,)^2\,\,,
\label{apacuatro}
\eeq
where $L^i$ is defined as:
\beq
L^i\,=\,2\,\tilde w^{\alpha}\,L_{\alpha}^i\,\,,
\label{apacinco}
\eeq
with $w^{i}$ being left invariant forms of the SU(2) group manifold.

We are interested in bosonic solutions of the equations of motion which are
supersymmetric. For this kind of solutions, the supersymmetric variations
of the
fermionic fields vanish for some Killing spinor $\epsilon$. In general, the
fermionic fields transform under supersymmetry as:
\bear
\delta\psi_{\lambda}&=&D_{\lambda}\,\epsilon\,+\,{1\over 24}\,e^{\phi}\,
F_{\mu\nu}^{i}\,\hat\Gamma_i\,(\,\Gamma_{\lambda}^{\mu\nu}\,-\,
10\,\delta_{\lambda}^{\mu}\,\Gamma^{\nu}\,)\,\epsilon\,-\,{1\over 288}\,
e^{-\phi}\epsilon_{ijk}\,\hat\Gamma^{ijk}\Gamma_{\lambda}\,T\epsilon\,,\rc\rc
\delta\chi_i&=&{1\over 2}\,(P_{\mu ij}\,+\,{2\over 3}\,\delta_{ij}\,
\partial_{\mu}\phi\,)\,\hat\Gamma^j\,\Gamma^{\mu}\,\epsilon\,-\,
{1\over 4}\,e^{\phi}\,F_{\mu\nu i}\,\Gamma^{\mu\nu}\,\epsilon\,-\,
{1\over 8}\,e^{-\phi}\,(\,T_{ij}\,-\,{1\over 2}\,\delta_{ij}\,T\,)\,
\epsilon^{jkl} \hat\Gamma_{kl}\epsilon\,,\rc\rc
\label{apaseis}
\eear
where the $\hat\Gamma_{i}$'s are the Dirac matrices along the SU(2) group
manifold and $D_{\mu}\,\epsilon$ is the covariant derivative  of the spinor
$\epsilon$, given by:
\beq
D_{\mu}\,\epsilon\,=\,\big(\,\partial_{\mu}\,+\,{1\over 4}\,
\omega_{\mu}^{ab}\,\Gamma_{ab}\,+\,{1\over 4}\,
Q_{\mu ij}\,\hat\Gamma^{ij}\,\big)\,\epsilon\,\,,
\label{apasiete}
\eeq
with $\omega_{\mu}^{ab}$ being the components of the spin connection.

\medskip
\renewcommand{\theequation}{\rm{B}.\arabic{equation}}
\setcounter{equation}{0}
\medskip
\section{Lagrangian approach to the round metric}
\medskip

In this appendix we are going to derive the first-order equations
(\ref{vuno}) by finding a superpotential for the effective lagrangian
${\cal L}_{eff}$ in eight dimensional supergravity. The first step in
this approach is to obtain the form of ${\cal L}_{eff}$ for the ansatz
given in eqs. (\ref{uno}) and (\ref{seis}). Actually, the expression of
${\cal L}_{eff}$ can be obtained by substituting (\ref{uno}) and (\ref{seis})
into the lagrangian given by eq. (\ref{apatres}). Indeed, one can check
that the
equations of motion of eight dimensional supergravity can be derived from the
following effective lagrangian:
\bear
{\cal L}_{eff}&=&e^{4f+3h}\,\Big[\,2(f'\,)^2\,+\,
(h'\,)^2\,-\,{1\over 3}\,(\phi'\,)^2\,-\,
4e^{2\phi-2h}\,(g'\,)^2\,+\,
4\,f'\,h'\,+\,\rc
&&+\,e^{-2h}\,+\,{1\over 16}\,e^{-2\phi}\,-\,
(4g^2\,-\,1)^2\,e^{2\phi-4h}\,\,\Big]\,\,,
\label{apbuno}
\eear
together with the zero-energy condition. In the equations obtained
from ${\cal L}_{eff}$ it is consistent to take $f=\phi/3$, which
we will do from now on. Next, let us introduce a new set of functions:
\beq
a\,=\,2\,e^{{2\phi\over 3}}\,\,,
\,\,\,\,\,\,\,\,\,\,\,\,\,\,\,
b\,=\,{1\over 2}\,e^{h\,-\,{\phi\over 3}}\,\,,
\label{apbdos}
\eeq
and a new variable $\eta$, defined as:
\beq
{dr\over d\eta}\,=\,e^{{4\phi\over 3}\,+\,3h}\,\,.
\label{apbtres}
\eeq
The effective lagrangian in these new variables has the kinetic term:
\beq
T\,=\,\Big(\,{\dot a\over a}\,\Big)^2\,+\,
\Big(\,{\dot b\over b}\,\Big)^2\,+\,3\,{\dot a\dot b\over a b}\,-\,
{1\over 4}\,{a^2\over b^2}\,\big(\,\dot g\,\big)^2\,\,,
\label{apbcuatro}
\eeq
where the dot denotes derivative with respect to $\eta$.
The potential in ${\cal L}_{eff}$ is:
\beq
V\,=\,{a^6b^6\over 2}\,\Big[\,(1\,-\,4g^2\,)^2\,
{a^2\over 32 b^4}\,-\,{1\over 2a^2}-\,{1\over 2b^2}\,\Big]\,\,.
\label{apbcinco}
\eeq
The superpotential for $T-V$ in the  variables just introduced has been
obtained in ref. \cite{CGLP1}, starting from eleven dimensional
supergravity. So,
we shall follow here the same steps as in ref. \cite{CGLP1} and
define $\alpha^1=\log a$,
$\alpha^2=\log b$ and $\alpha^3=\log g$. Then, the kinetic energy $T$ can be
rewritten as:
\beq
T\,=\,{1\over 2}\,g_{ij}\,{d\alpha^i\over d\eta}\,
{d\alpha^j\over d\eta}\,\,,
\label{apbseis}
\eeq
where $g_{ij}$ is the matrix:
\beq
g_{ij}\,=\,\pmatrix{2&&3&&0\cr
                    3&&2&&0\cr
                    0&&0&&-{a^2\over 2b^2}}\,\,.
\label{apbsiete}
\eeq
The superpotential $W$ for this system must satisfy:
\beq
V\,=\,-{1\over 2}\,g^{ij}\,
{\partial W\over \partial \alpha^i}\,
{\partial W\over \partial \alpha^j}\,\,,
\label{apbocho}
\eeq
where $g^{ij}$ is the inverse of $g_{ij}$ and $V$ has been written in eq.
(\ref{apbcinco}). By using the values of $g_{ij}$  in (\ref{apbsiete}),
one can write explicitly the relation between $V$ and $W$ as:
\beq
V\,=\,{1\over 5}\,a^2\,\Bigg({\partial W\over \partial a}\Bigg)^2\,+\,
{1\over 5}\,b^2\,\Bigg({\partial W\over \partial b}\Bigg)^2\,-\,
{3\over 5}\,ab\,{\partial W\over \partial a}\,{\partial W\over 
\partial b}\,+\,
{b^2\over a^2}\,\Bigg({\partial W\over \partial g}\Bigg)^2\,\,.
\label{apbnueve}
\eeq
Moreover, it is not difficult to verify, following again ref. \cite{CGLP1},
that
$W$ can be taken as:
\beq
W\,=\,{1\over 8}\,a^2\,b\,
\sqrt{\Big(\,a^2\,(1-2g)^2\,+\,4b^2\,\Big)\,
\Big(\,a^2\,(1+2g)^2\,+\,4b^2\,\Big)}\,\,.
\label{apbdiez}
\eeq
The first-order equations associated to the superpotential $W$ are:
\beq
{d\alpha^i\over d\eta}\,=\,g^{ij}\,{\partial W\over \partial\alpha^j}\,\,.
\label{apbonce}
\eeq
By substituting the expressions of $W$ and $g^{ij}$ on the right-hand side of
eq. (\ref{apbonce}), and by writing the result in terms of the
variables used in section 2, one can check that the system (\ref{apbonce})
is the same as that written in eq. (\ref{vuno}).

\end{document}